\documentclass[printer]{aa} 
\usepackage{graphicx}
\usepackage{lscape}
\usepackage{txfonts}
\begin{document}
   \title{Deep surface photometry of edge-on spirals in Abell galaxy clusters:}

   \subtitle{constraining environmental effects}

   \author{B. X. Santiago
          \inst{1}
          \and
          T. B. Vale\inst{1}
          }

   \offprints{B. Santiago}

   \institute{Instituto de F\'\i sica, Universidade Federal do Rio Grande\\ 
              do Sul, Av. Bento Gon\c calves, 9500, CP 15051, Porto Alegre\\
              \email{santiago@if.ufrgs.br}}

   \date{Received September 15, 1996; accepted March 16, 1997}

 
  \abstract
   {There is a clear scarcity of structural parameters for stellar thick discs, 
   especially for spiral galaxies located in high-density regions, such as 
   galaxy clusters and compact groups.}
   {We have modelled the thin and thick discs of 4 edge-on 
   spirals located in Abell clusters: NGC 705, ESO243G49, ESO187G19, 
   LCSBS0496P. Deep I band images of NGC 705 were taken from the HST 
   archive, whereas the remaining images were obtained with the 
   Southern Telescope for Astrophysical Research (SOAR) in Gunn r filter. 
   They reached surface brightness levels of $\mu_I \simeq 26.0$ mag 
   arcsec$^{-2}$ and $\mu_r \simeq 26.5$ mag arcsec$^{-2}$, respectively.}
   {Profiles were extracted from the deep images, in directions both parallel 
   and perpendicular to the major axis. Profile fits were carried
   out at several positions, yielding horizontal and vertical scale parameters
   for both thin and thick disc components.}
   {The extracted profiles and fitted disc parameters vary from galaxy to 
   galaxy. Two galaxies have a horizontal profile with a strong
   down-turn at outer radii, preventing a simple exponential from fitting
   the entire range. For the 2 early-type 
   spirals, the thick discs have larger scalelengths than the thin discs,
   whereas no trend is seen for the later types . Both 
   the thin and thick discs sampled tend to have similar scalelengths and 
   scaleheights when compared to typical field disc galaxies. However, the 
   thin disc parameters of the 2 farthest galaxies, both late-type spirals, 
   may be significantly affected by seeing effects. Taken at face value, our
   results suggest that environment plays a minor role in determining the 
   thin and thick disc sizes.}
   {}


   \authorrunning{Santiago \& Vale}

   \titlerunning{Edge-on spirals in Abell Clusters}

   \maketitle
%

\section{Introduction}

At high surface brightness, the light of disc galaxies is dominated by the
central spheroidal bulge and the planar disc. The structure and the 
relative contributions of these two components make up the basis of the
morphological classification system and drive our knowledge of
galactic structure. In our own Galaxy, however, most of the 
information on the  early processes of mass assembly and 
star formation is hidden
in the unconspicuous and low surface-brightness components, such as
the Galactic halo and the thick disc.

The Milky Way's thick disc has been studied in much more detail than
in any other galaxy. Its scaleheight is about 3 times that of the old
thin disc, and it is formed by old ($\tau > 10 Gyrs$) and has moderately
metal-poor ($[Fe/H \simeq -1.0$) stars with intermediate kinematical
properties between the thin disc and the stellar halo (Wyse 1999, 
Norris 1999, Kerber, Javiel \& Santiago 2001, Du et al. 2003).
Data on external thick discs and haloes are hard to obtain, because of the
extremely low intensity levels. As a consequence, relatively few and
largely nearby galaxies have been studied in depth enough to sample
these earlier structural components. 

In a few nearby galaxies, the extended thick disc has
been studied using resolved stellar population techniques (Mould 2005, Tikhonov
, Galazutdinova \& Drozdovsky 2005, Seth, Dalcanton \& de Jong 2005). 
Seth et al. (2005)
used the Hubble Space Telescope Advanced Camera for Surveys
in order to analyse the vertical distribution of stars in low-mass galaxies.
They find stellar components well above the thin disc scaleheight and
properties similar to the Milky Way thick disc. They also find
an increasing age as a function of height consistent with disc heating.

However, most galaxies are still too far to be resolved in stars. 
This means that their extraplanar light can only be analysed with surface 
photometry work applied on deep CCD images (Morrison, Boroson \& Harding 
1994, van Dokkum et al. 1994, Morrison et al. 1997, Abe et al. 1999, 
Fry et al. 1999, Neeser et al. 2002, Wu et al. 2002). 
Edge-on orientation is certainly favourable in order to distinguish 
the vertically extended components from the thin disc and 
central bulge.
Several photometric surveys of edge-on spirals have been undertaken, 
yielding thin disc structural parameters and, at best, allowing
detection of thick disc light (de Grijs 1998,
Dalcanton \& Bernstein 2000,2002). However, these surveys were not deep
enough to allow determination of thick disc structural parameters.
Taken together, these studies have revealed that most galaxies
exhibit extraplanar light that may often be identified with a thick-disc 
component. Most of them have demonstrated that 
a single exponential, or a single $sech^2$ surface brightness profile, 
is not an 
adequate fit to the vertical distribution of light. Instead, a second
component is required, usually with a uniform scaleheight and a larger 
scalelength as compared to the thin disc.
Another interesting development has been the realization that the
disc structure parallel to the major axis may often require a 
2-piecewise exponential fit, where the change in exponential scale
occurs at some (anti-)truncation radius (Pohlen et al. 2004, 2007).

One important feature of most studies aiming at either detecting or
measuring the thick disc component in unresolved galaxies is that
these tend to be nearby objects (so that the disc scales are resolved well
and more easily measured) inhabiting the general field.
Environmental effects may have played an important role in shaping the
extended low surface-brightness components in galaxies, just as they
have done with bulges and thin discs. This may be especially true in
scenarios with thick disc formation through accretion events
in a hierarchical structure formation 
such as the one proposed by Abadi et al. (2003) and Brook et al. (2004).

This paper is an initial effort to systematically address 
a possible environmental effect on thick disc structure by 
analysing deep images of 4 edge-on spirals 
located in Abell clusters. Once a sizeable
sample is available for different high-density environments,
our final goal is to statistically assess the environmental dependence 
of the structural
parameters of extra-planar light in spirals of different types and masses.


\section{Data and photometry}

   The sample was chosen through inspection of Digital Sky Survey (DSS) 
   fields containing rich Abell clusters in the sourthern equatorial 
   hemisphere. Whenever an edge-on candidate was found, we searched for
   information on it in the NASA Extragalactic Database (NED). 
   Our list
   includes only objects with semi-major axis $a >
   20''$, axis ratio $b/a < 0.3$ and with a measured 
   redshift whose value is close to that of the Abell cluster. Whenever
   possible the Hubble Space Telescope (HST) archive was searched 
   in order to confirm the edge-on orientation of the candidate in
   higher resolution . The galaxies studied here are listed in 
   Table \ref{sample}, where we
   provide the basic information on them.

   \begin{table}
      \caption[]{The sample.}
         \label{sample}
     $$ 
         \begin{array}{p{0.18\linewidth}p{0.12\linewidth}p{0.15\linewidth}p{0.15\linewidth}p{0.09\linewidth}p{0.07\linewidth}p{0.07\linewidth}p{0.05\linewidth}}
            \hline
            \noalign{\smallskip}
            Galaxy  &  Abell Cluster & RA$^a$ & Dec.$^a$ & Type &
            SMA$^a$ ($''$) & b/a$^a$ & z$^a$ \\
            \noalign{\smallskip}
            \hline
            \noalign{\smallskip}
            NGC705 & A0262 & 01:52:41.5 & +36:08:38 & S0/a & 36 & 0.21 & 0.016 \\
            ESO243G49 & A2877 & 01:10:27.7 & -46:04:27 & S0/a & 32 & 0.28 & 0.022 \\
            LCSB0496P & A0419 & 03:09:20.3 & -23:45:00 & Late S & 22 & 0.18 & 0.041\\
            ESO187G19 & A3716 & 20:51:06.3 & -52:42:45 & Sb & 50 & 0.12 & 0.047 \\
            \noalign{\smallskip}
            \hline
         \end{array}
     $$
\begin{list}{}{}
\item[$^{\mathrm{a}}$] Source: positions, sizes and redshifts were taken from NED. SMA in column 6 stands for semi-major axis. $b/a$ in column 7 is the axis ratio.
\end{list}
   \end{table}

   \begin{table}
      \caption[]{SOAR/SOI Observation logs.}
         \label{logs}
     $$ 
         \begin{array}{p{0.2\linewidth}p{0.1\linewidth}p{0.1\linewidth}p{0.1\linewidth}}
            \hline
            \noalign{\smallskip}
            Galaxy  &  Filter & $N_{exp}$ & Exp. Time (s)\\
            \noalign{\smallskip}
            \hline
            \noalign{\smallskip}
            ESO243G49 & Gunn r & 14 & 720 \\
            ESO187G19 & Gunn r & 15 & 720 \\
            LCSBS0496P & Gunn r & 14 & 720 \\
            \noalign{\smallskip}
            \hline
         \end{array}
     $$ 
   \end{table}

   For NGC 705, archival images in the F814W filter
   were found to be deep enough to allow the analysis without 
   further imaging. 
   A total of 11 raw individual exposures totalling 12700s = 3.5hrs 
   were retrieved, reduced, and stacked as discussed below.

   Images of ESO243G49, ESO187G19, and LCSBS0496P were obtained in late 2005
   with the Optical Imager (SOI) at the Southern Telescope for 
   Astrophysical Research (SOAR). Seeing was always with $FWHM < 1.1''$.
   The SOI has 2 E2V CCDs,
   each one with 2k x 4k pixels, covering a field of view of 5.5X5.5 
   arcminutes. We used the 2x2 binning, yielding a detector scale of 0.154
   arcsec/pixel. To reduce instrumental noise, the images were 
   taken on the slow readout mode. The log of observations is given in 
   Table \ref{logs}. A total of about 3hrs exposure was taken in the Gunn r
   filter for each galaxy. As our aim was to measure the disc
   structural parameters, and also to test the SOI detection limits at 
   low surface
   brightness levels, we decided to make deep exposures on a single filter
   sensitive to old stellar populations rather than to acquire colour 
   information.

\begin{figure}
   \centering
   \begin{tabular}{cc}
   \includegraphics[width=0.25\textwidth]{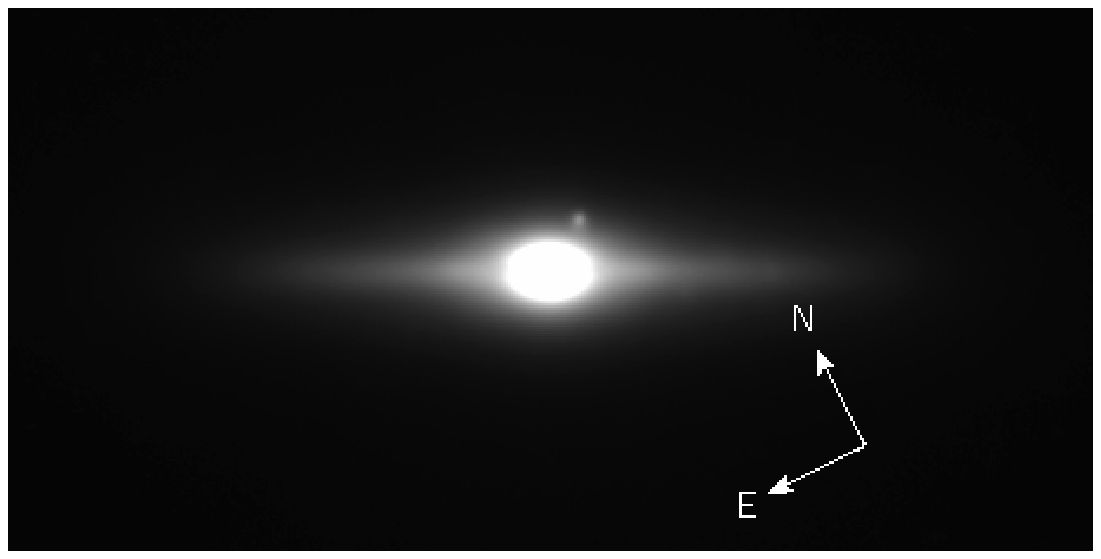}
   \includegraphics[width=0.25\textwidth]{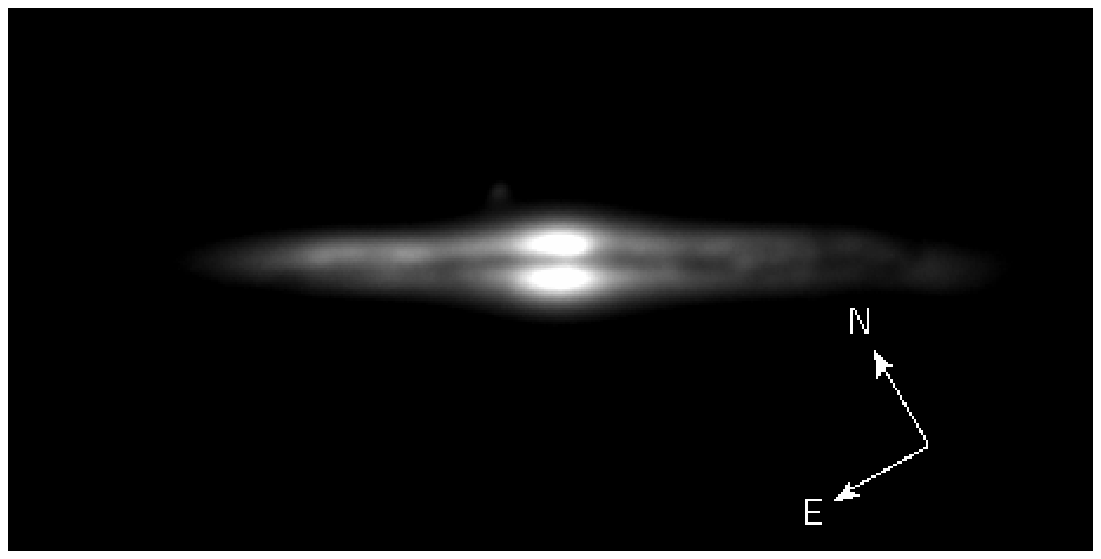} \\
   \includegraphics[width=0.25\textwidth]{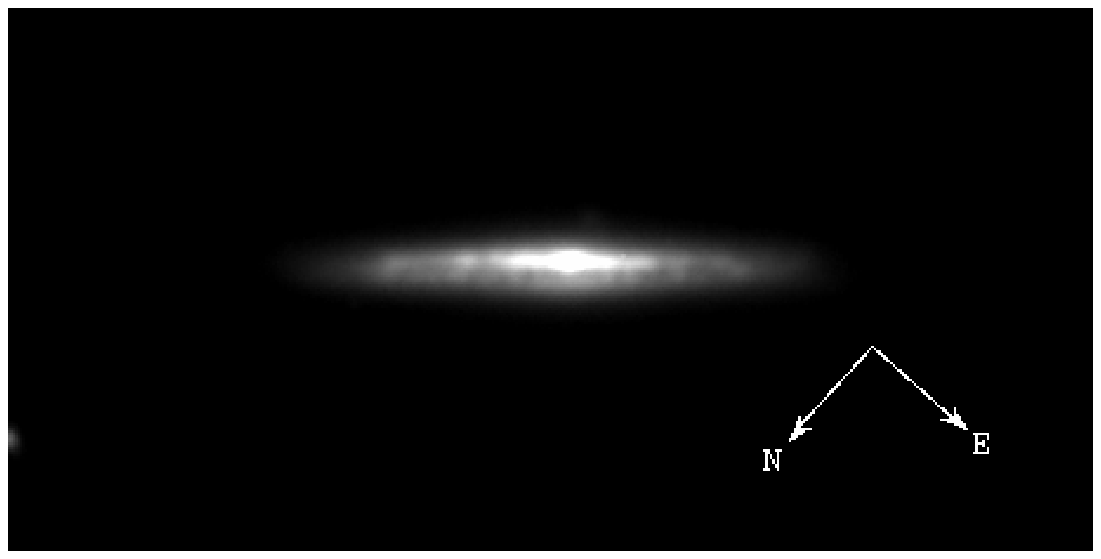}
   \includegraphics[width=0.25\textwidth]{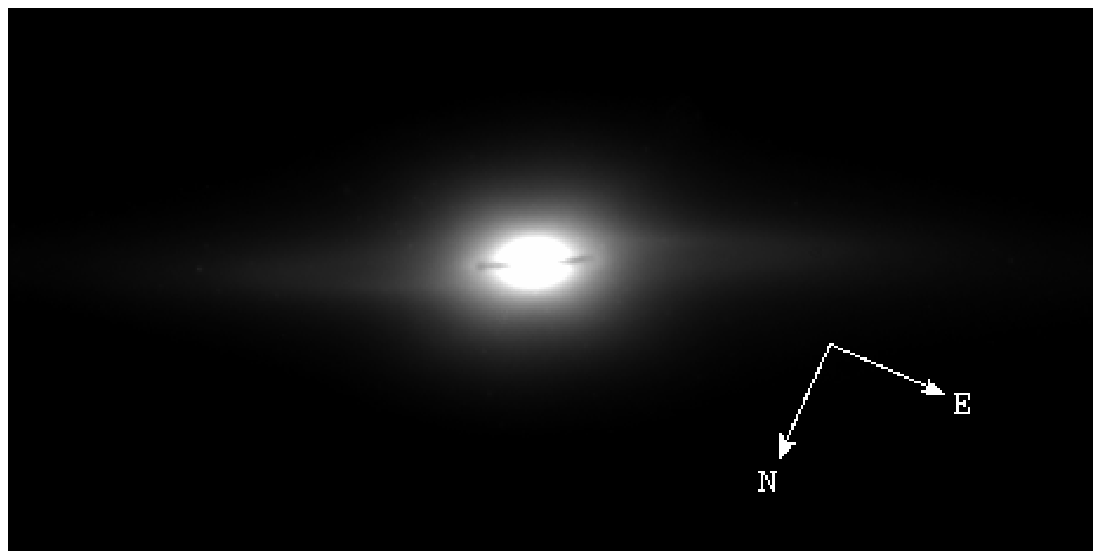} \\
   \end{tabular}
   \caption{Final images of the 4 cluster galaxies studied in this
    paper; upper-left: ESO243G49; upper-right: ESO187G19; lower-left: 
    LCSBS0496P; lower-right: NGC 705. All panels are shown with the
    same contrast levels}
    \label{soiimgs}
    \end{figure}

\begin{figure}
   \centering
   \begin{tabular}{cc}
   \includegraphics[width=0.25\textwidth]{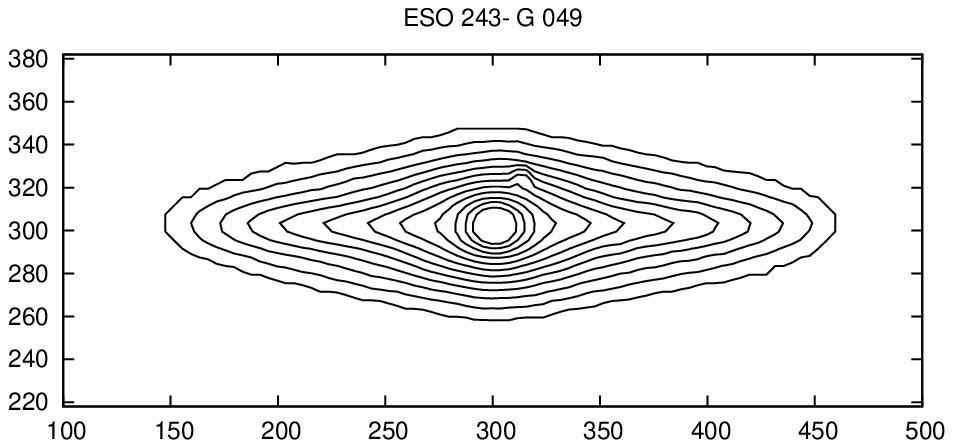}
   \includegraphics[width=0.25\textwidth]{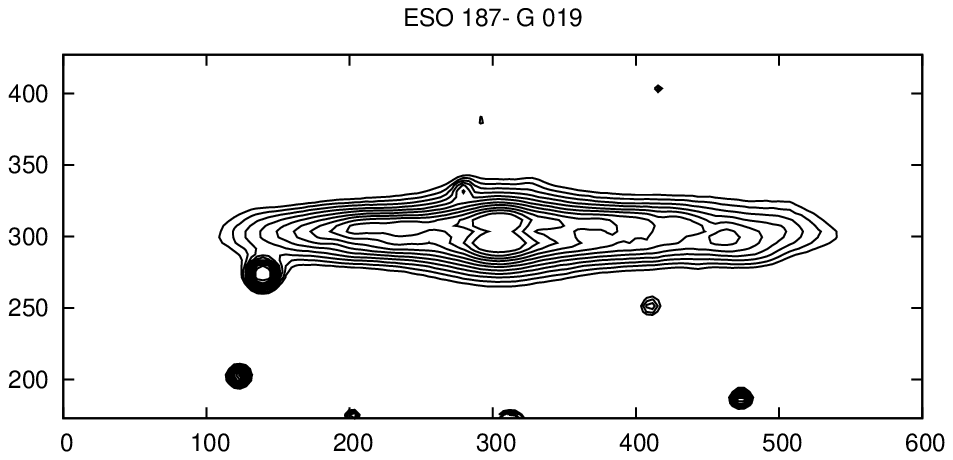} \\
   \includegraphics[width=0.25\textwidth]{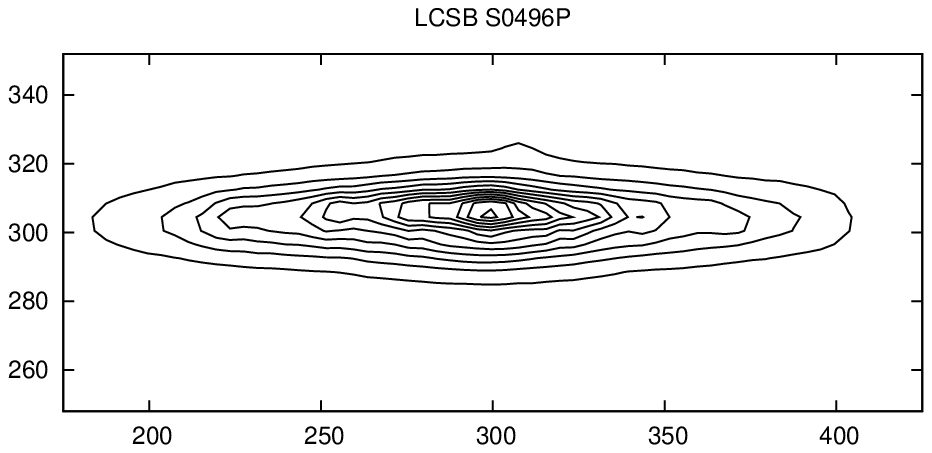}
   \includegraphics[width=0.25\textwidth]{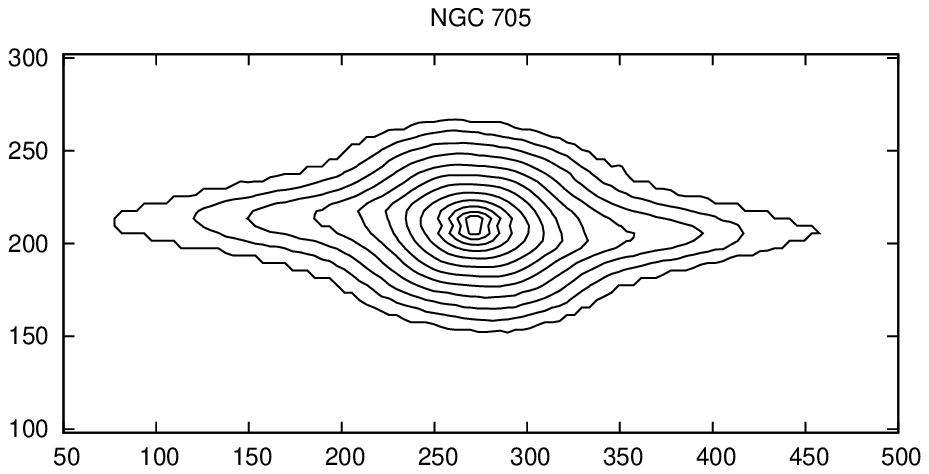} \\
   \end{tabular}
   \caption{Contour plots of the galaxies in the sample. Each panel
     has 12 isophotal levels fitted in log-scale}
    \label{soicont}
    \end{figure}

   The images were trimmed, bias-subtracted, and flat-fielded using high 
   S/N dome flats. The individual exposures were combined taking the 
   positional dithering into account. The
   final combined images are shown in Figure \ref{soiimgs}. Contour
   plots of the same galaxies are shown in Figure \ref{soicont}.

   The raw WFPC2/HST frames of NGC 705 were retrieved from the 
   archive, pipelined using IRAF stsdas.wfpc package, and combined. 
   A total of 11 F814W raw frames were successfuly retrieved 
   and stacked in the process. Its final image is also shown in 
   Figure \ref{soiimgs}. 

   \begin{figure}
   \centering
   \includegraphics[width=0.5\textwidth]{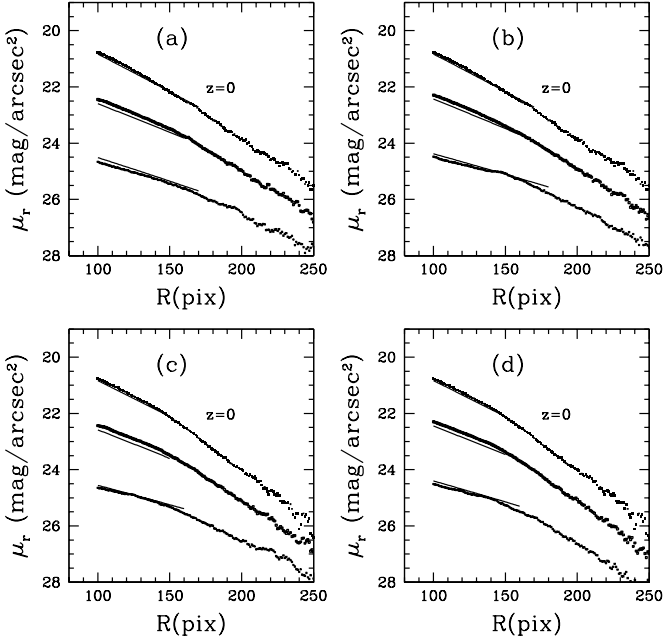}
   \caption{Surface brightness profiles of ESO243G49 in directions 
               parallel to its major axis. The profiles are ordered 
               according to increasing $z$ value, as indicated
               in Table \ref{eso243rtab}. Except for the uppermost 
               profile (major axis,$z=0$), a vertical shift was 
               applied to avoid cluttering. The shift size was 
               simply $\Delta \mu_r = n$ mag arcsec$^{-2}$, where $n$ 
               is the integer listed in the 5$^{th}$ column of the 
               mentioned table. The solid lines are fits to the data, 
               as described in the text.}
              \label{eso243r}%
    \end{figure}

   \begin{table}
      \caption[]{ESO243G49 $R$ profile regions.}
         \label{eso243rtab}
     $$
         \begin{array}{p{0.30\linewidth}p{0.13\linewidth}p{0.22\linewidth}p{0.07\linewidth}p{0.08\linewidth}p{0.08\linewidth}p{0.08\linewidth}}
            \hline
            \noalign{\smallskip}
         Extract. Range & Central pixel & Fit Range & panel & n & $R_0$ & $R_0$ \\
         ($z_{min}$:$z_{max}$,$R_{min}$:$R_{max}$) & & (min:max) & & & pix & $h_{75}^{-1}$ kpc \\
            \noalign{\smallskip}
            \hline
            \noalign{\smallskip}
         (-4:+4,100:260) & z=0 & (100:170) & a,b & 0 & 35.9 & 2.5 \\
         (-20:-5,100:260) & z=-12.5 & (100:160) & a & 1 & 46.0 & 3.2 \\
         (-44:-21,100:260) & z=-32.5 & (100:170) & a & 2 & 54.6 & 3.8 \\
         (+5:+20,100:260) & z=+12.5 & (100:160) & b & 1 & 44.5 & 3.1 \\
         (21:+44,100:260) & z=+32.5 & (100:180) & b & 2 & 61.8 & 4.3 \\
         (-4:+4,-260:-100) & z=0 & (-150:-100) & c,d & 0 & 37.4 & 2.6 \\
         (-20:-5,-260:-100) & z=-12.5 & (-150:-100) & c & 1 & 46.0 & 3.2 \\
         (-44:-21,-260:-100) & z=-32.5 & (-160:-100) & c & 2 & 64.7 & 4.5 \\
         (+5:+20,-260:-100) & z=+12.5 & (-150:-100) & d & 1 & 46.0 & 3.2 \\
         (21:+44,-260:-100) & z=+32.5 & (-160:-100) & d & 2 & 60.3 & 4.2 \\
         \noalign{\smallskip}
         \hline
         \end{array}
     $$
\begin{list}{}{}
\item [] Notes: Column 1: Range perpendicular (zmin:zmax) and parallel 
(Rmin:Rmax) to major axis where the profile was extracted, in pixels. Pixel
scale is 0.154arcsec/pixel; column 2: central pixel position, in pixels; 
column 3: fit range in pixels used; column 4: panel in 
Figure \ref{eso243r} where the
profile is shown; column 5: artificial shift in surface brightness 
applied to profile in its panel, $\Delta \mu_r = n$ mag arcsec$^2$.
The larger $n$, the lower the profile position in its panel; column 6:
fitted scalength, $R_0$, in pixels; column 7: scalelength in
units of $h_{75}^{-1}~kpc$, where $h_{75} = H_0 / 75$ and 
$H_0$ is the Hubble constant in units of km s$^{-1}$ Mpc$^{-1}$.
\end{list}
   \end{table}

   Two very important steps were applied to the combined images: 
   sky-subtraction and masking of contaminating sources. Sky subtraction
   was performed by iteratively fitting a background plane over points 
   equally spaced on a CCD grid. Grid points that deviated by more 
   than $2\sigma$ from the median value were rejected, and the median 
   and $\sigma$ were recomputed until convergence. This $\sigma$ 
   clipping procedure prevented stars and the 
   target galaxies themselves from contaminating 
   the sky fit. After sky 
   subtraction we cut out an image section of the SOI or WFPC2 chip, 
   which 
   conservatively contained  the entire galaxy image down to its lowest 
   detectable levels. This image section was then searched for faint 
   sources, either foreground and background, which were masked out 
   by replacing their counts with the mean count within a ring 
   around them. The ring size was varied manually so as to allow
   an effective masking of contaminated pixels. The search for and
   masking of contaminating sources was carried out only in regions 
   that were later to be used for the structural modelling.
   As the galaxies
   are not superposed on crowded stellar fields, the fraction of
   masked pixels in these regions was always below 20\%.

   As a final step before the analysis, the images were rotated 
   to align the galaxy major axis parallel to the horizontal 
   image borders. As common in this type of study, the alignement 
   procedure was done with the fainter and outer isophotes.
   
   Since our goal is to detect the disc components and measure their 
   horizontal and vertical scales, strict photometric calibration was not
   necessary. For SOI/SOAR, a zero point to the Gunn $\it{r}$ magnitude scale 
   was found by measuring the instrumental magnitudes and comparing them to
   the Cousins R magnitudes available in NED. The $R_T$ values are quoted in
   NED for ESO243G49 and ESO187G19. Aperture photometry within a circle 
   of radius $r < 50''$ with
   a zero-point of 25.35 recovers their quoted magnitudes with residuals of 
   0.02 and -0.04, respectively. We applied the same zero-point to the
   third galaxy imaged with SOAR, LCSBS0496P. For NGC 705, we used the 
   photometric zero-point quoted by Holtzmann et al. (1995) for WFPC2.

\section{Profile fits}

   Profiles both parallel and perpendicular to each galaxy's major axis
   were extracted from the final processed image. We hereafter
   refer to the image coordinates that run parallel (perpendicular) to
   the major axis as $R$ ($z$). The extraction 
   windows are listed in Tables \ref{eso243rtab} thru \ref{ngcztab}.
   All these tables have the same structure. 
   Column 1 lists the extraction windows [zmin:zmax,Rmin:Rmax] in
   pixels. Column 2 gives the central pixel along which the profile
   was extracted. Column 3 gives the range in pixels used in fitting
   the profiles. In the case of the $z$ profiles, two regions are
   indicated, one for fitting the thin disc, and the outer one
   where the thick disc was fitted. Column 4 indicates the figure
   panel and column 5 gives the position (from top to bottom) in 
   this panel where the profile is plotted.
   Finally, the last two columns list the fit parameters. For the
   horizontal profiles, only a single exponential fit was carried out,
   yielding a scalelength $R_0$, whereas the $z$ profiles fits led 
   to the two scaleheights, of
   the thin and thick disc components ($z_{01}$ and $z_{02}$, 
   respectively). The profile parameters are shown both in pixels
   and in units of $h_{75}^{-1}$ kpc, where $h_{75} = H_0 / 75$ and
   $H_0$ is the Hubble constant in units of km s$^{-1}$ Mpc$^{-1}$.
   For simplicity, we will not explicitely use the Hubble constant
   parameterization throughout the text when referring to physical scales.
   Details on the profile fits are given below. 

   \begin{figure}
   \centering
   \includegraphics[width=0.5\textwidth]{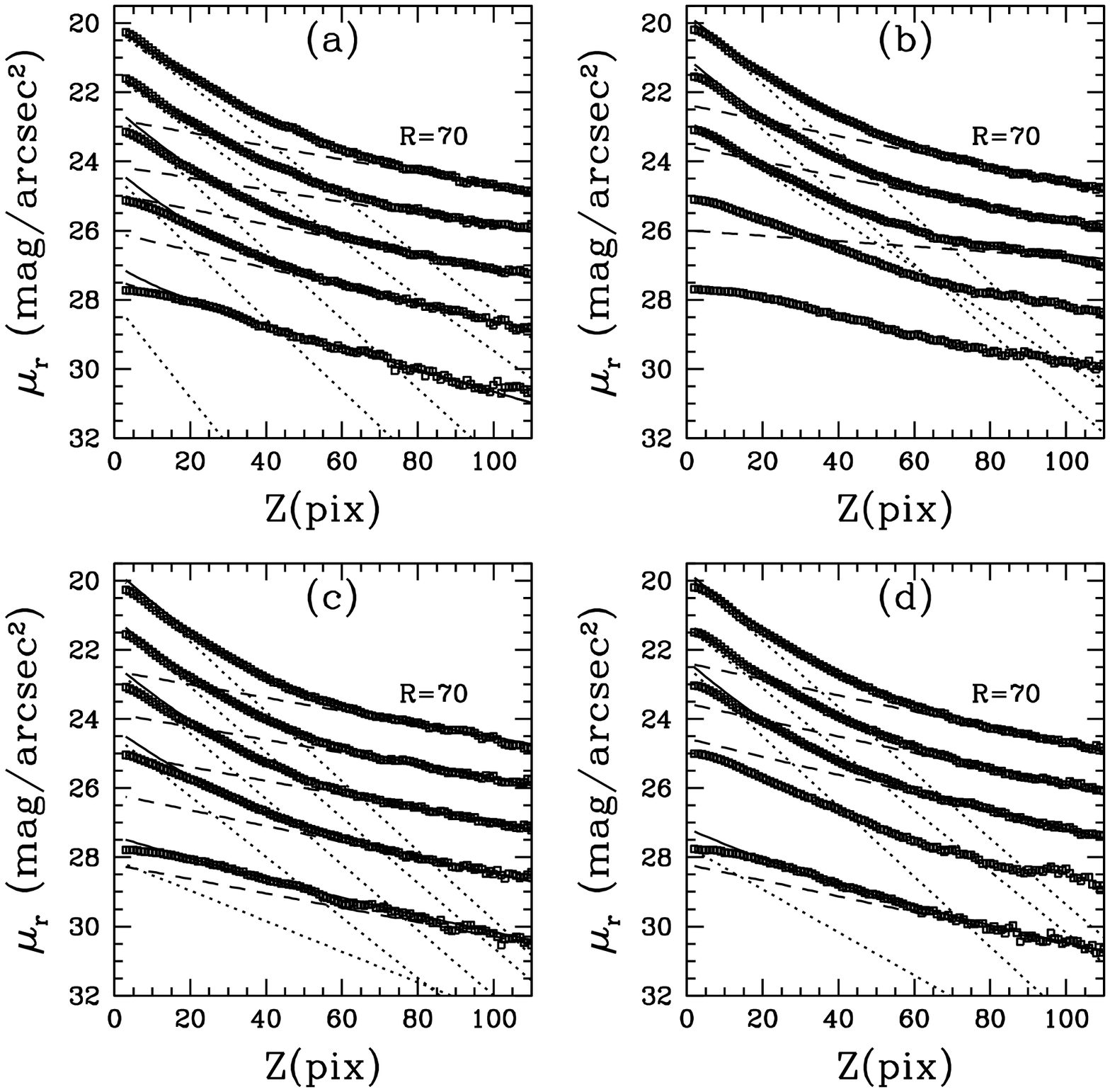}
   \caption{Surface brightness profiles of ESO243G49 in directions 
               perpendicular to its major axis. The profiles are ordered 
               according to increasing $R$ value, as indicated
               in Table \ref{eso243ztab}. Except for the uppermost 
               profile ($R=\pm 70pix$), a vertical shift was 
               applied to avoid cluttering. The shift size was 
               simply $\Delta \mu_r = n$ mag arcsec$^{-2}$, where $n$ 
               is the integer listed in the 5$^{th}$ column 
               of the mentioned table. The dotted (dashed) line is the 
               fitted exponential
               thin (thick) disc. The total thin+thick disc contribution
               is given by the solid lines.}
              \label{eso243z}%
    \end{figure}

   \begin{table}
      \caption[]{ESO243G49 $z$ profile regions.}
         \label{eso243ztab}
     $$
         \begin{array}{p{0.28\linewidth}p{0.15\linewidth}p{0.28\linewidth}p{0.06\linewidth}p{0.05\linewidth}p{0.08\linewidth}p{0.08\linewidth}}
            \hline
            \noalign{\smallskip}
         Extract. Range & Central pixel & Fit Range & pnl & n & $z_{01}$, $z_{02}$ & $z_{01}$, $z_{02}$ \\
         ($z_{min}$:$z_{max}$,$R_{min}$:$R_{max}$) & & (min:max) & & & pix & $h_{75}^{-1}$ kpc \\
            \noalign{\smallskip}
            \hline
            \noalign{\smallskip}
(-120:-3,60:80) & R=70 & Thin:(-50:-20), Thick:(-100:-60) & a & 0 & 13.4, 56.0 & 0.9, 3.9 \\
(3:120,60:80) & R=70 & Thin:(20:50), Thick:(60:100) & b & 0 & 11.4, 48.0 & 0.8, 3.3 \\ 
(-120:-3,80:100) & R=90 & Thin:(-50:-20), Thick:(-100:-60) & a & 1 & 13.6, 62.6 & 0.9, 4.4 \\
(3:120,80:100) & R=90 & Thin:(20:50), Thick:(60:100) & b & 1 & 11.1, 49.1 & 0.8, 3.4 \\
(-120:-3,100:130) & R=115 & Thin:(-50:-20), Thick:(-100:-60) & a & 2 & 10.8, 49.6 & 0.7, 3.5 \\
(3:120,100:130) & R=115 & Thin:(20:50), Thick:(60:100) & b & 2 & 15.6, 139.8 & 1.1, 9.7 \\
(-120:-3,130:170) & R=150 & Thin:(-50:-20), Thick:(-100:-60) & a & 3 & 10.5, 42.5 & 0.7, 2.9 \\
(3:120,130:170) & R=150 & Thin:(20:50), Thick:(60:100) & b & 3 & NA, NA & NA, NA \\
(-120:-3,170:230) & R=200 & Thin:(-50:-20), Thick:(-100:-60) & a & 4 & 7.9, 33.7 & 0.6, 2.3 \\
(3:120,170:230) & R=200 & Thin:(20:50), Thick:(60:100) & b & 4 & NA, NA & NA, NA \\
(-120:-3,-80:-60) & R=-70 & Thin:(-50:-20), Thick:(-100:-60) & c & 0 & 10.8, 56.4  & 0.7, 3.9 \\
(3:120,-80:-60) & R=-70 & Thin:(20:50), Thick:(60:100) & d & 0 & 11.4, 46.5 & 0.8, 3.2 \\
(-120:-3,-100:-80) & R=-90 & Thin:(-50:-20), Thick:(-100:-60) & c & 1 & 11.5, 57.7 & 0.8, 4.0 \\
(3:120,-100:-80) & R=-90 & Thin:(20:50), Thick:(60:100) & d & 1 & 12.3, 45.5 & 0.9, 3.2 \\
(-120:-3,-130:-100) & R=-115 & Thin:(-50:-20), Thick:(-100:-60) & c & 2 & 11.5, 53.9 & 0.8, 3.7 \\
(3:120,-130:-100) & R=-115 & Thin:(20:50), Thick:(60:100) & d & 2 & 10.7, 41.5 & 0.7, 2.9 \\
(-120:-3,-170:-130) & R=-150 & Thin:(-50:-20), Thick:(-100:-60) & c & 3 & 12.4, 47.4 & 0.9, 3.3 \\
(3:120,-170:-130) & R=-150 & Thin:(20:50), Thick:(60:100) & d & 3 & NA, NA & NA, NA \\
(-120:-3,-230:-170) & R=-200 & Thin:(-50:-20), Thick:(-100:-60) & c & 4 & 24.5, 52.3 & 1.71, 3.6 \\
(3:120,-230:-170) & R=-200 & Thin:(20:50), Thick:(60:100) & d & 4 & 17.5, 47.1 & 1.2, 3.3 \\
            \noalign{\smallskip}
            \hline
         \end{array}
     $$
\begin{list}{}{}
\item [] Notes: Column 1: Range perpendicular (zmin:zmax) and parallel 
(Rmin:Rmax) to major axis where the profile was extracted, in pixels. Pixel
scale is 0.154arcsec/pixel; column 2: central pixel position, in pixels; 
column 3: fit range, in pixels, used for the thin and thick components; 
column 4: panel in Figure \ref{eso243z} where 
the profile is shown; column 5: artificial shift in surface brightness 
applied to profile in its panel, $\Delta \mu_r = n$ mag arcsec$^2$.
The larger $n$, the lower the profile position in its panel; column 6:
fitted scaleheights, $z_{01}$ and $z_{02}$, in pixels; column 7: 
scaleheights in units of $h_{75}^{-1}~kpc$, where $h_{75} = H_0 / 75$ and 
$H_0$ is the Hubble constant in units of km s$^{-1}$ Mpc$^{-1}$.
\end{list}
   \end{table}

   The $R$ and $z$ profiles are shown in Figures \ref{eso243r} thru
   \ref{ngc705z}. The panels shown in these figures correspond to the 
   different quadrants relative to the centre and major axis of each
   galaxy. The quadrants are identified in the figures using the same
   panel labels as in Tables \ref{eso243rtab} thru \ref{ngcztab}. The 
   orientation of the profiles in Figures \ref{soiimgs} and \ref{soicont}
   is such that $z$ increases upwards and $R$ increases towards the
   left in the images.

   The extraction and fit regions were chosen
   to make use of as much independent information as possible
   from the images, especially at the faint surface
   brightness levels. Thus, for the $z$ profiles, the extraction range 
   in $R$ was increased
   at larger distances from the centre to increase the 
   signal-to-noise ratio (S/N). Similarly, wider ranges in $z$
   were used to extract the $R$ profiles far from the major axis. For
   a fixed profile, we chose not to bin the profile data unequally in  
   order to increase the signal at the faintest levels, as this would
   weight the number of data points against these lowest intensity 
   levels, where we expect the
   thick disc to dominate. This procedure, of course, does not lead to a 
   minimized effect of random noise on the fits, but we opted to pay this
   price in order to better sample the thick disc component.

   \begin{figure}
   \centering
   \includegraphics[width=0.5\textwidth]{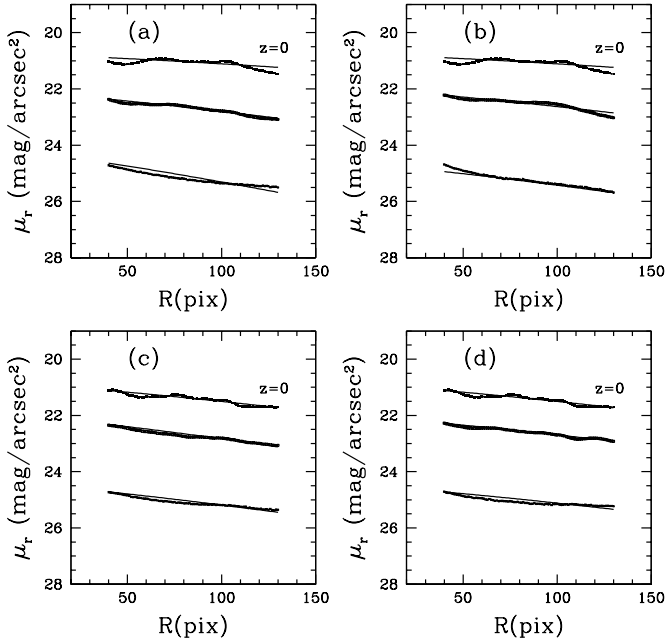}
   \caption{Surface brightness profiles of ESO187G19 in directions 
               parallel to its major axis. The profiles are ordered 
               according to increasing $z$ value, as indicated
               in Table \ref{eso187rtab}. Except for the uppermost 
               profile (major axis,$z=0$), a vertical shift was 
               applied to avoid cluttering. The shift size was 
               simply $\Delta \mu_r = n$ mag arcsec$^{-2}$, where $n$ 
               is the integer listed in the 5$^{th}$ column of the 
               mentioned table. The solid lines are fits to the data, 
               as described in the text.}
              \label{eso187r}%
    \end{figure}

   The fitting function applied to the $R$ profiles was 

   \begin{equation}
      I(R,z) = I(0,z)~\frac{R}{R_0}~K_1(\frac{R}{R_0}) 
   \end{equation}

   \noindent{
   where $I$ is the intensity level at image coordinates $(R,z)$,
   $I(0,z)$ is the central intensity for that profile, $R_0$ is the
   horizontal exponential scale and $K_1$ is the modified Bessel
   function of the second type (van der Kruit \& Searle 1981). 
   The fit was carried out using
   $R_0$ and $I(0,z)$ as free parameters in each case. 
}
   We adopted an exponential function for the $z$ profiles:

   \begin{equation}
      I(R,z) = I(R,0)~e^{-z/z_0}
   \end{equation}

\begin{figure}
   \centering
   \includegraphics[width=0.5\textwidth]{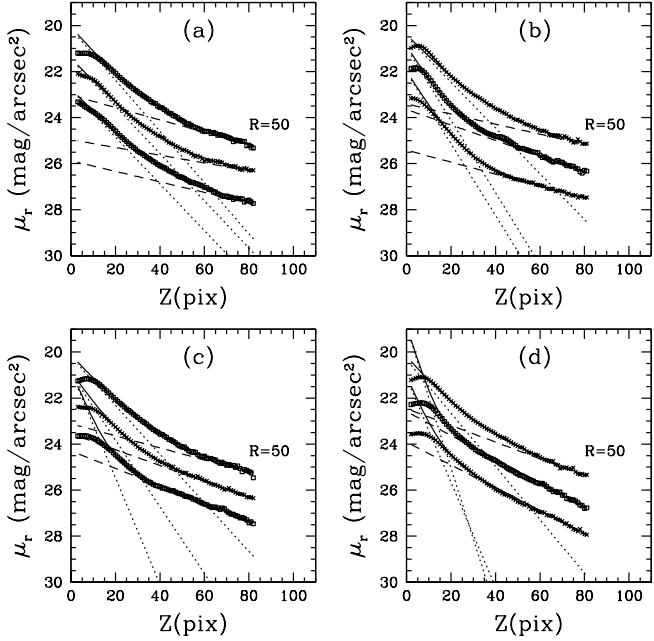}
   \caption{Surface brightness profiles of ESO187G19 in directions 
               perpendicular to its major axis. The profiles are ordered 
               according to increasing $R$ value, as indicated
               in Table \ref{eso187ztab}. Except for the uppermost 
               profile ($R=\pm 50$ pix), a vertical shift was 
               applied to avoid cluttering. The shift size was 
               simply $\Delta \mu_r = n$ mag arcsec$^{-2}$, where $n$ 
               is the integer listed in the 5$^{th}$ column of the 
               mentioned table. The dotted (dashed) line is the fitted 
               exponential thin (thick) disc. The total thin+thick disc 
               contribution is given by the solid lines.}
              \label{eso187z}%
    \end{figure}

   \noindent{
   for each one of the disc components. In the equation above $I(R,0)$
   is the intensity at the galaxy disc mid-plane at the 
   each profile position and $z_0$ the exponential scaleheight.
}
   We assumed in the fits that the two discs are co-planar. The fits
   therefore have 4 free parameters, namely, the mid-plane intensities
   and the scaleheights. The $z$ profile fits were done in an iterative
   way: a first fit to the function shown in equation (2) 
   was made in the outer $z$ regions, 
   where the thick disc dominates. The
   fitted thick disc was then extrapolated towards the disc plane and
   subtracted from the observed profile within the inner regions listed
   in the profile tables. The residuals were then fitted to the thin disc,
   and the result was then extrapolated outwards. The procedure was
   repeated until convergence to within 5\% was reached for all 4
   parameters. For some vertical profiles, the fits did not converge 
   after 10 iterations or led to unphysical solutions. We still
   plot these profiles in Figures \ref{eso243z}, \ref{eso187z}, 
   \ref{lcsbz}, and \ref{ngc705z} but 
   flag them in the parameter tables with an {\it NA}. 

   We now separately present the profiles and fitting results for 
   each galaxy.

\subsection{ESO243G49}

   This galaxy is an early type spiral located at about 10 arcmin 
   ($\simeq 0.3$ Mpc in projected separation)
   from IC1633, the central dominant (cD) galaxy of Abell 2877. 
   It has essentially the same redshift as the cluster (Abell 2877:
   $z = 0.023$; ESO243G49: $z = 0.022$). The $R$ profiles shown in 
   Figure \ref{eso243r} follow the same shape with little
   variation from one quadrant to the next. Such symmetric appearance
   shows that ESO243G49 has little dust extinction and is fairly 
   undisturbed.
   The symmetry also indicates that ESO243G49 is very close to a perfect
   edge-on orientation. 

   In each panel in the figure, the upper $R$ profiles are those along 
   the major axis, whereas
   the other two are at positions parallel to it. 
   They systematically reach surface brightness levels of $\mu_r \simeq 
   26.5$ mag arcsec$^{-2}$. Notice that the inner
   profile region shown is always 100 pixels ($\simeq 7$ kpc) 
   or farther away from the centre,
   to avoid the bright bulge (clearly visible in Figure \ref{soiimgs}). 
   Also note that all these profiles have a break or down-bending
   in the outer regions (beyond $R \simeq 170$ pix $\simeq 14$ kpc); 
   this feature is not accounted for by the fits to an 
   exponential light distribution with an edge-on projection. This
   down-bending has
   been previously reported on other edge-on spirals 
   (see Pohlen et al. 2004, 2007) and is discussed later.

   \begin{table}
      \caption[]{ESO187G19 $R$ profile regions.}
         \label{eso187rtab}
     $$
         \begin{array}{p{0.30\linewidth}p{0.13\linewidth}p{0.22\linewidth}p{0.07\linewidth}p{0.08\linewidth}p{0.08\linewidth}p{0.08\linewidth}}
            \hline
            \noalign{\smallskip}
         Extract. Range & Central pixel & Fit Range & panel & n & $R_0$ & $R_0$ \\
         ($z_{min}$:$z_{max}$,$R_{min}$:$R_{max}$) & & (min:max) & & & pix & $h_{75}^{-1}$ kpc \\
            \noalign{\smallskip}
            \hline
            \noalign{\smallskip}
(-4:+4,40:130) & z=0 & (40:130) & a,b & 0 & 175.1 & 22.3 \\
(-19:-5,40:130) & z=-12.0 & (40:130) & a & 1 & 100.5 & 12.8 \\
(-41:-20,40:130) & z=-30.5 & (40:130) & a & 2 & 70.7 & 9.0 \\
(+5:+19,40:130) & z=+12.0 & (40:130) & b & 1 & 107.5 & 13.7 \\
(20:+41,40:130) & z=+30.5 & (40:130) & b & 2 & 92.6 & 11.8 \\
(-4:+4,-130:-40) & z=0 & (-130:-40) & c,d & 0 & 109.9 & 14.0 \\
(-19:-5,-130:-40) & z=-12.0 & (-130:-40) & c & 1 & 87.9 & 11.2 \\
(-41:-20,-130:-40) & z=-30.5 & (-130:-40) & c & 2 & 95.0 & 12.1 \\
(+5:+19,-130:-40) & z=+12.0 & (-130:-40) & d & 1 & 106.0 & 13.5 \\
(20:+41,-130:-40) & z=+30.5 & (-130:-40) & d & 2 & 108.3 & 13.8 \\
            \noalign{\smallskip}
            \hline
         \end{array}
     $$
\begin{list}{}{}
\item [] Notes: Column 1: Range perpendicular (zmin:zmax) and parallel 
(Rmin:Rmax) to major axis where the profile was extracted, in pixels. Pixel
scale is 0.154arcsec/pixel; column 2: central pixel position, in pixels; 
column 3: fit range in pixels used; column 4: panel in 
Figure \ref{eso187r} where the
profile is shown; column 5: artificial shift in surface brightness 
applied to profile in its panel, $\Delta \mu_r = n$ mag arcsec$^2$.
The larger $n$, the lower the profile position in its panel; column 6:
fitted scalength, $R_0$, in pixels; column 7: scalelength in
units of $h_{75}^{-1}~kpc$, where $h_{75} = H_0 / 75$ and 
$H_0$ is the Hubble constant in units of km s$^{-1}$ Mpc$^{-1}$.
\end{list}
   \end{table}

   The fits were carried out at the inner regions, before the 
   observed break in the profiles, and are also shown in the panels. 
   As mentioned
   earlier, the resulting scalelengths, $R_0$, are listed in Table
   \ref{eso243rtab}. A straight average over the individual profile
   fits yields $R_0 = 3.5 \pm 0.7$ kpc, with no strong deviant points.
   Even though the values scatter by no more than 20\% around the mean,
   it is worth noticing the trend toward increasing $R_0$ for the profiles 
   farther away from the disc plane. This may be the result of a larger
   thick disc scalelength as compared to the thin disc. If we take the
   last profiles of each panel in Figure \ref{eso243r} as representative
   of the thick-disc horizontal light distribution, we have an $R_0 = 4.2 \pm 
   0.3$ for this component. This is 1.6 times greater than the scalelength
   resulting from the two major axis profiles.

   In Figure \ref{eso243z} we show the $z$ profiles for ESO243G49. Again,
   each panel refers to the profiles in a particular quadrant of the 
   galaxy image. The symmetry among the corresponding profiles in each 
   panel is again clearly visible. Also, nearly all these profiles 
   deviate significantly from a straight line. This means that a 
   single exponential
   disc model cannot account for the profile shapes. A two-component
   model is therefore necessary to successfully fit the data. These two-
   component fits are shown in the panels and were carried out as described
   in the beginning of this section. The fits avoid the 
   20 pixels (3.2arcsec) 
   closest to the plane of the galaxy, where seeing effects may be 
   important. 

   As a rule, these resulting fits from a thin and a thick disc are a very
   good description of the observed profiles. In general, the fitted 
   scaleheights vary little from one profile to the other. For the thin
   disc, we have $z_{01} = 0.9 \pm 0.3$ kpc when we average all profiles.
   Eliminating one obvious outlier yields $z_{01} = 0.8 \pm 0.1$ kpc,
   the same result as applying an iterative $2\sigma$ clipping algorithm 
   to the original 
   distribution of $z_{01}$. The average thick-disc scaleheight is
   $z_{02} = 3.8 \pm 1.6$ kpc. Again, the values listed in Table 
   \ref{eso243ztab} indicate a clearly deviating fitting result. If 
   eliminated, the spread around the average is reduced a lot, yielding
   $z_{02} = 3.4 \pm 0.5$ kpc. A formal $2\sigma$ clipping yields 
   $z_{02} = 3.5 \pm 0.4$ kpc.

\subsection{ESO187G19}

   \begin{table}
      \caption[]{ESO187G19 $z$ profile regions.}
         \label{eso187ztab}
     $$
         \begin{array}{p{0.28\linewidth}p{0.15\linewidth}p{0.28\linewidth}p{0.06\linewidth}p{0.05\linewidth}p{0.08\linewidth}p{0.08\linewidth}}
            \hline
            \noalign{\smallskip}
         Extract. Range & Central pixel & Fit Range & pnl & n & $z_{01}$, $z_{02}$ & $z_{01}$, $z_{02}$ \\
         ($z_{min}$:$z_{max}$,$R_{min}$:$R_{max}$) & & (min:max) & & & pix & $h_{75}^{-1}$ kpc \\
            \noalign{\smallskip}
            \hline
            \noalign{\smallskip}
(-80:-3,40:60) & R=50 & Thin:(-60:-15), Thick:(-80:-60) & a & 0 & 9.8, 40.6 & 1.4, 5.9 \\
(3:80,40:60) & R=50 & Thin:(15:40), Thick:(60:80) & b & 0 & 10.9, 49.4 & 1.6, 7.2 \\ 
(-80:-3,60:90) & R=75 & Thin:(-60:-15), Thick:(-80:-60) & a & 1 & 10.6, 64.6 & 1.5, 9.4 \\
(3:80,60:90) & R=75 & Thin:(15:35), Thick:(60:80) & b & 1 & 6.9, 32.3 & 1.0, 4.7 \\
(-80:-3,90:140) & R=115 & Thin:(-60:-15), Thick:(-80:-60) & a & 2 & 10.7, 47.5 & 1.6, 6.9 \\
(3:80,90:140) & R=115 & Thin:(15:35), Thick:(60:80) & b & 2 & 7.0, 41.8 & 1.0, 6.1 \\
(-80:-3,-60:-40) & R=-50 & Thin:(-45:-15), Thick:(-80:-60) & c & 0 & 10.3, 41.5 & 1.5, 6.1 \\
(3:80,-60:-40) & R=-50 & Thin:(15:40), Thick:(60:80) & d & 0 & 9.4, 30.0 & 1.4, 4.4 \\
(-80:-3,-90:-60) & R=-75 & Thin:(-40:-15), Thick:(-80:-60) & c & 1 & 7.5, 31.5 & 1.1, 4.6 \\
(3:80,-90:-60) & R=-75 & Thin:(15:40), Thick:(60:80) & d & 1 & 3.5, 21.0 & 0.5, 3.1 \\
(-80:-3,-140:-90) & R=-115 & Thin:(-35:-15), Thick:(-80:-60) & c & 2 & 4.7, 28.8 & 0.7, 4.2 \\
(3:80,-140:-90) & R=-115 & Thin:(15:40), Thick:(60:80) & d & 2 & 4.8, 21.9 & 0.7, 3.2 \\
            \noalign{\smallskip}
            \hline
         \end{array}
     $$
\begin{list}{}{}
\item [] Notes: Column 1: Range perpendicular (zmin:zmax) and parallel 
(Rmin:Rmax) to major axis where the profile was extracted, in pixels. Pixel
scale is 0.154arcsec/pixel; column 2: central pixel position, in pixels; 
column 3: fit range, in pixels, used for the thin and thick components; 
column 4: panel in Figure \ref{eso187z} where the 
profile is shown; column 5: artificial shift in surface brightness 
applied to profile in its panel, $\Delta \mu_r = n$ mag arcsec$^2$.
The larger $n$, the lower the profile position in its panel; column 6:
fitted scaleheights, $z_{01}$ and $z_{02}$, in pixels; column 7: 
scaleheights in units of $h_{75}^{-1}~kpc$, where $h_{75} = H_0 / 75$ and 
$H_0$ is the Hubble constant in units of km s$^{-1}$ Mpc$^{-1}$.
\end{list}
   \end{table}

   ESO187G19 is a member of Abell 3716 and is located some 2 arcmin 
   ($\simeq 0.1$ Mpc in projected separation) from
   the cluster centre, although its measured redshift, $z=0.047$, 
   is a bit higher than that of the cluster ($z=0.041$). 
   In many respects it is a very distinct galaxy when compared to 
   ESO243G49. It is a latter type spiral, with a weak spheroid visible
   only in its very central regions. It displays a strong dust lane, which
   cuts the central bulge through its centre. Assuming that this dust 
   absorption originates from the edge of an optically thick thin disc,
   the inclination of ESO187G19 should be $i=90^o$. The galaxy disc
   is clearly flared in its outer areas ($R > 130$ pixels $= 20.8$
   arcsec $\simeq 17$ kpc), so our profiles are truncated
   at this radius, and the fits also avoid these flared regions. 
   On the other hand, the much weaker
   bulge allows the inner fit region to be $R > 40$ pix $= 6.4$ arcsec.

   The resulting horizontal profiles and fits are shown in
   Figure \ref{eso187r}. The uppermost ones, corresponding to the 
   major axis,
   are obviously affected by dust, displaying a wavy shape, not
   adequately fit by an exponential. The model fits for these profiles 
   are shown just for completeness. The profile distortions are very
   reduced in the remaining $R$ profiles. The corresponding profiles
   in each quadrant are again very similar, showing the symmetry
   in the light distribution. Also,
   the profiles run almost parallel to each other, with no sign of
   the break in the outer regions seen in the case of ESO243G49. This 
   difference is likely to be real, since in physical units the 
   fitting regions used are very similar in both galaxies.
   
   Averaging over all fits, we find a scalelength $R_0 = 13.4 \pm 
   3.5$ kpc.
   This large scatter is largely caused by a single deviant point, with
   $R_0 = 22.3$ kpc. Notice that this profile runs along the major 
   axis and is strongly affected by the dust lane, as just mentioned. 
   By applying a $2\sigma$ clipping to
   eliminate this outlier, a much more precise $R_0 = 12.9 \pm 1.0$ kpc
   is obtained, using 8 out of the 10 profiles shown.
   In contrast to ESO243G49, no clear trend toward increasing $R_0$ is 
   seen at greater distances from the disc plane.

   \begin{figure}
   \centering
   \includegraphics[width=0.5\textwidth]{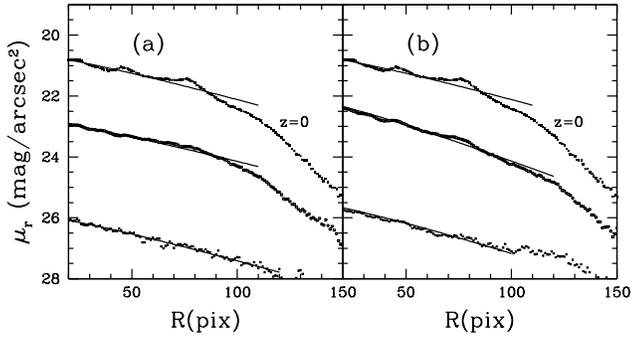}
   \caption{Surface brightness profiles of LCSBS0496P in directions 
               parallel to its major axis. The profiles are ordered 
               according to increasing $z$ value, as indicated
               in Table \ref{lcsbrtab}. Except for the uppermost 
               profile (major axis,$z=0$), a vertical shift was 
               applied to avoid cluttering. The shift size was 
               simply $\Delta \mu_r = n$ mag arcsec$^{-2}$, where $n$ 
               is the integer listed in the 5$^{th}$ column of the 
               mentioned table. The solid lines are fits to the data, 
               as described in the text.}
              \label{lcsbr}%
    \end{figure}

   The vertical profiles of ESO187G19 are shown in Figure
   \ref{eso187z}. Their
   interpretation is certainly complicated by the clear dust lane, which 
   causes the leveling or down-turning of the profiles at low $z$.
   The effect of seeing, given the much greater distance, also 
   severely affects the inner region. Still, a cut in $z = 15$ pix 
   $= 2.4$ arcsec from the mid-plane corresponds to 
   2.5-3.0 times the seeing FWHM, so that the light spread from the
   galaxy midplane is strongly reduced. In physical terms, this
   cut corresponds to $z > 1.9$ kpc, which is well within the region 
   dominated by the thick disc in our Galaxy. It is therefore quite 
   noticeable that, 
   even considering this cut-off in $z$, the profiles have a similar
   shape to those observed in ESO243G49, at about half the distance. The
   $z$ profiles cannot be adequately described by a single straight 
   line, even beyond the inner cut-off radius. 
   Instead, they still show two distinct vertical ranges where 
   such lines can be fit.
   Therefore, although being cautious about the resulting thin
   disc scaleheight, we still fitted a double
   exponential to the observed profiles and show the results 
   in the figure.

   \begin{table}
      \caption[]{LCSBS0496P $R$ profile regions.}
         \label{lcsbrtab}
     $$
         \begin{array}{p{0.30\linewidth}p{0.13\linewidth}p{0.22\linewidth}p{0.07\linewidth}p{0.08\linewidth}p{0.08\linewidth}p{0.08\linewidth}}
            \hline
            \noalign{\smallskip}
         Extract. Range & Central pixel & Fit Range & panel & n & $R_0$ & $R_0$ \\
         ($z_{min}$:$z_{max}$,$R_{min}$:$R_{max}$) & & (min:max) & & & pix & $h_{75}^{-1}$ kpc \\
            \noalign{\smallskip}
            \hline
            \noalign{\smallskip}
(-4:+4,20:150) & z=0 & (20:110) & a,b & 0 & 49.5 & 6.3 \\
(-17:-5,20:150) & z=-11.0 & (20:110) & a & 1 & 52.6 & 6.7 \\
(-32:-18,20:150) & z=-25.0 & (20:120) & a & 2 & 48.7 & 6.2 \\
(+5:+17,20:150) & z=+11.0 & (20:120) & b & 1 & 37.7 & 4.8 \\
(18:+32,20:150) & z=+25.0 & (20:100) & b & 2 & 44.0 & 5.6 \\
            \noalign{\smallskip}
            \hline
         \end{array}
     $$
\begin{list}{}{}
\item [] Notes: Column 1: Range perpendicular (zmin:zmax) and parallel 
(Rmin:Rmax) to major axis where the profile was extracted, in pixels. Pixel
scale is 0.154arcsec/pixel; column 2: central pixel position, in pixels; 
column 3: fit range in pixels used; column 4: panel in 
Figure \ref{lcsbr} where the 
profile is shown; column 5: artificial shift in surface brightness 
applied to profile in its panel, $\Delta \mu_r = n$ mag arcsec$^2$.
The larger $n$, the lower the profile position in its panel; column 6:
fitted scalength, $R_0$, in pixels; column 7: scalelength in
units of $h_{75}^{-1}~kpc$, where $h_{75} = H_0 / 75$ and 
$H_0$ is the Hubble constant in units of km s$^{-1}$ Mpc$^{-1}$.
\end{list}
   \end{table}

   The profiles were fit out to $\mu_r \simeq 26.0$ mag
   arcsec$^{-2}$. Within the adopted fitting range, the joint 
   contribution of thin and thick disc provide a 
   very efficient description of the observed profiles.
   The fitted parameters are not as stable as in the 
   case of ESO243G49.
   This is naturally reflected in the dispersion around the average
   scaleheights. For the thin disc we obtain $z_{01} = 1.2 \pm 0.4$ 
   kpc, with no values exceeding $2\sigma$ from the mean in either
   direction. The thick disc fit results also show substantial
   spread around the average. If all points are considered we 
   derive $z_{02} = 5.5 \pm 1.8$ kpc. Applying a formal $2\sigma$ clipping
   results in $z_{02} = 5.1 \pm 1.4$ kpc. 

\subsection{LCSBS0496P}

   In contrast to the other galaxies studied in this paper, LCSBS0496P
   is not close to the centre of its cluster. Its projected 
   distance from the centre of Abell 0419 is about 13 arcmin 
   ($\simeq 0.6$ Mpc). It is therefore found
   in relative isolation and has perhaps not had time to fall into
   the more crowded regions of Abell 0419. On the other hand,
   we note that its measured redshift is exactly the same as 
   that attributed to Abell 0419. 

   \begin{figure}
   \centering
   \includegraphics[width=0.5\textwidth]{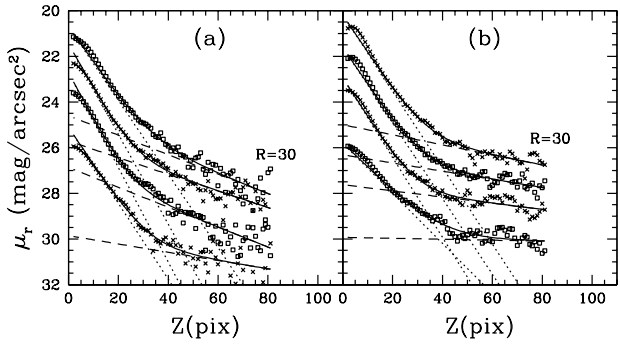}
   \caption{Surface brightness profiles of LCSBS0496P in directions 
               perpendicular to its major axis. The profiles are ordered 
               according to increasing $R$ value, as indicated
               in Table \ref{lcsbztab}. Except for the uppermost 
               profile ($R=30$ pix), a vertical shift was 
               applied to avoid cluttering. The shift size was 
               simply $\Delta \mu_r = n$ mag arcsec$^{-2}$, where $n$ 
               is the integer listed in the 5$^{th}$ column of the 
               mentioned table. The dotted (dashed) line is the fitted 
               exponential thin (thick) disc. The total thin+thick disc 
               contribution is given by the solid lines.}
              \label{lcsbz}%
    \end{figure}

   This galaxy is at a comparable distance to 
   ESO187G19. It is also
   a late type, rather bulgeless, spiral at an inclination close 
   to $i=90^o$. This conclusion again comes from the assumption 
   that the aborption lane is generated at the edge of an
   optically thick thin disc. As a result of its very late type,
   profiles could be extracted at $R > 20$ pix from its centre.

   The $R$ profiles of LCSBS0496P are shown in Figure \ref{lcsbr},
   reaching down at least to $\mu_r \simeq 26.0$ mag arcsec$^{-2}$. 
   Notice that only the left part of the image shown in 
   Figure \ref{soiimgs} ($R>0$) was used in the fits. The reason is that
   the regions on the right were contaminated by some residual 
   scattered light 
   that often affected the borders of the SOI images at its
   early-science stage. From the remaining profiles, we obtain
   $R_0 = 5.9 \pm 0.7$ kpc with no outliers.

   The profiles exhibit a
   behaviour similar to that of the early-type ESO243G49. Most of the 
   profiles shown in the figure present an abrupt downturn at an
   outer radius of $R \simeq 110$ pix $= 14$ kpc, 
   preventing a single exponential model from being
   applied to the entire profile range. On the other hand, there is 
   no clear trend toward increasing the scalelength towards the region
   where the thick disc dominates.

   The vertical profiles shown in Figure \ref{lcsbz} are also
   affected by dust and seeing at the very centre, as in the case of
   ESO187G19. The $z$ profiles are much noisier and less symmetric than in the
   previous two galaxies. LCSBS0496P also seems to display a
   steeper profile perpendicular to the disc plane, reaching the
   sky level at smaller $z$. This significantly shortens the
   available $z$ range for fitting the two disc components, which
   clearly reflects onto the resulting parameters. Most especially, the
   thick disc scaleheight ($z_{02}$) varies considerably from one
   profile to another. Because of the greater 
   difficulty in fitting two disc components to such shorter and noisier data,
   we relaxed the parameter convergence criteria to $10\%$, rather than $5\%$,
   in our iterative fitting algorithm. We also had to push the fitting ranges
   closer to the disc plane in some cases.

   For the thin disc, an average $z_{01} = 0.8 \pm 0.2$ results
   from the fits. Applying a $2\sigma$ clipping eliminates one outlier
   and reduces the scatter but does not alter the scaleheight: $z_{01} = 
   0.8 \pm 0.1$. This is $30\%$ smaller than in the case of
   ESO187G19. In contrast to ESO187G19, there is very little variation
   in the thin disc fit results among the different profiles. 
   As LCSBS0496P is
   also at a large distance, we are confident that we are seeing
   real physical variations in scaleheights from one galaxy to
   another, even though the scaleheights may be systematically
   affected by seeing. On the other hand, the thick disc has been
   much less constrained by our fits, which result in a broad
   distribution of scaleheights, with a mean of $z_{02} = 
   5.4 \pm 2.3$ kpc. This large uncertainty accommodates the 6 
   individual estimates.

   \begin{table}
      \caption[]{LCSBS0496P $z$ profile regions.}
         \label{lcsbztab}
     $$
         \begin{array}{p{0.28\linewidth}p{0.15\linewidth}p{0.28\linewidth}p{0.06\linewidth}p{0.05\linewidth}p{0.08\linewidth}p{0.08\linewidth}}
            \hline
            \noalign{\smallskip}
         Extract. Range & Central pixel & Fit Range & pnl & n & $z_{01}$, $z_{02}$ & $z_{01}$, $z_{02}$ \\
         ($z_{min}$:$z_{max}$,$R_{min}$:$R_{max}$) & & (min:max) & & & pix & $h_{75}^{-1}$ kpc \\
            \noalign{\smallskip}
            \hline
            \noalign{\smallskip}
(-80:-3,20:40) & R=30 & Thin:(-35:-15), Thick:(-80:-40) & a & 0 & 6.5, 24.0 & 0.8, 3.1 \\
(3:80,20:40) & R=30 & Thin:(15:30), Thick:(30:80) & b & 0 & 6.4, 44.9 & 0.8, 5.7 \\ 
(-80:-3,40:60) & R=50 & Thin:(-25:-10), Thick:(-80:-30) & a & 1 & 5.4, 28.7 & 0.7, 3.6 \\
(3:80,40:60) & R=50 & Thin:(15:35), Thick:(35:80) & b & 1 & 6.1, 49.8 & 0.8, 6.3 \\
(-80:-3,60:90) & R=75 & Thin:(-25:-10), Thick:(-80:-30) & a & 2 & 5.2, 24.4 & 0.7, 3.1 \\
(3:80,60:90) & R=75 & Thin:(15:30), Thick:(35:80) & b & 2 & 6.1, 72.2 & 0.8, 9.2 \\
(-80:-3,90:140) & R=115 & Thin:(-30:-10), Thick:(-60:-35) & a & 3 & 6.4, 51.9 & 0.8, 6.6 \\
(3:80,90:140) & R=115 & Thin:(15:30), Thick:(30:80) & b & 3 & 9.5, 168.1 & 1.2, 21.4 \\
            \noalign{\smallskip}
            \hline
         \end{array}
     $$
\begin{list}{}{}
\item [] Notes: Column 1: Range perpendicular (zmin:zmax) and parallel 
(Rmin:Rmax) to major axis where the profile was extracted, in pixels. Pixel
scale is 0.154arcsec/pixel; column 2: central pixel position, in pixels; 
column 3: fit range, in pixels, used for the thin and thick components; 
column 4: panel in Figure \ref{lcsbz} where the 
profile is shown; column 5: artificial shift in surface brightness 
applied to profile in its panel, $\Delta \mu_r = n$ mag arcsec$^2$.
The larger $n$, the lower the profile position in its panel; column 6:
fitted scaleheights, $z_{01}$ and $z_{02}$, in pixels; column 7: 
scaleheights in units of $h_{75}^{-1}~kpc$, where $h_{75} = H_0 / 75$ and 
$H_0$ is the Hubble constant in units of km s$^{-1}$ Mpc$^{-1}$.
\end{list}
   \end{table}

\subsection{NGC 705}

\begin{figure}
   \centering
   \includegraphics[width=0.5\textwidth]{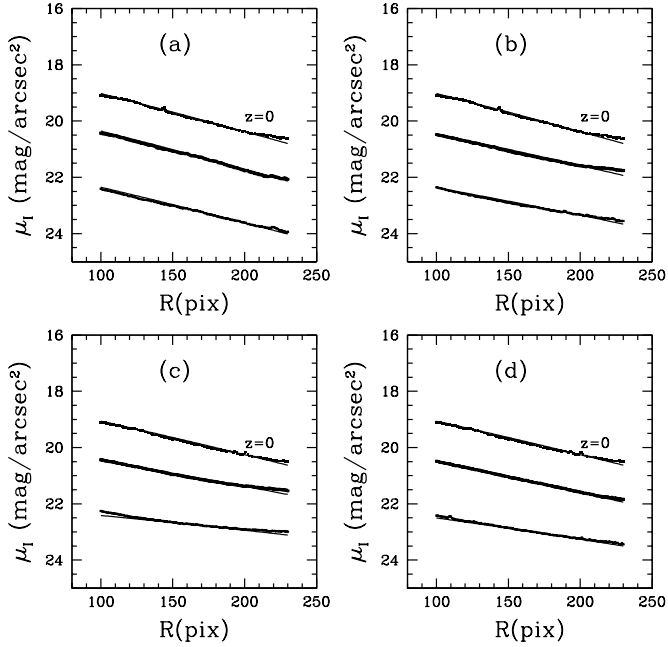}
   \caption{Surface brightness profiles of NGC 705 in directions 
               parallel to its major axis. The profiles are ordered 
               according to increasing $z$ value, as indicated
               in Table \ref{ngcrtab}. Except for the uppermost 
               profile (major axis,$z=0$), a vertical shift was 
               applied to avoid cluttering. The shift size was 
               simply $\Delta \mu_I = n$ mag arcsec$^{-2}$, where $n$ 
               is the integer listed in the 5$^{th}$ column of the 
               mentioned table. The solid lines are fits to the data, as 
               described in the text.}
              \label{ngc705r}%
    \end{figure}

\begin{figure}
   \centering
   \includegraphics[width=0.5\textwidth]{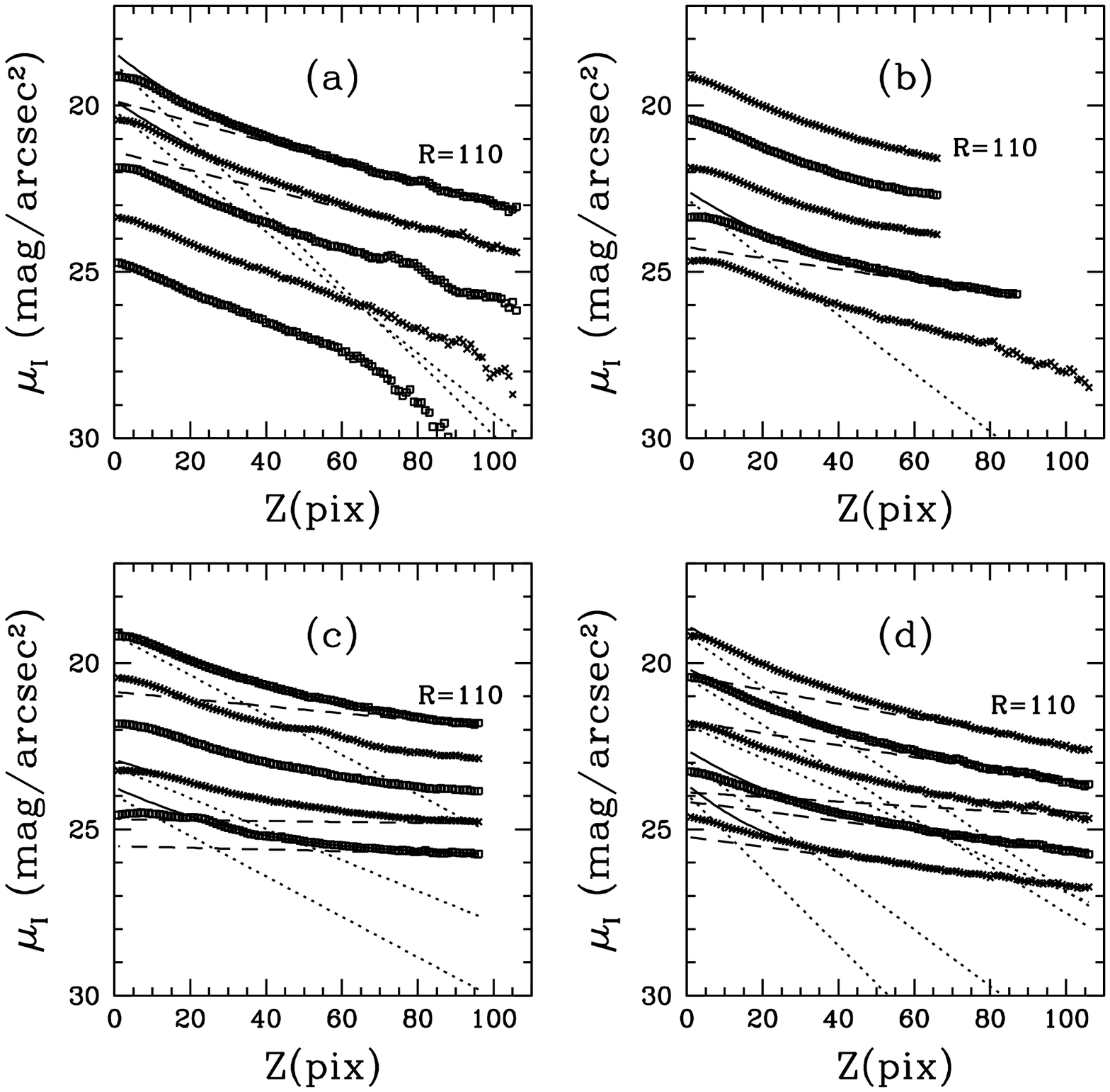}
   \caption{Surface brightness profiles of NGC 705 in directions 
               perpendicular to its major axis. The profiles are ordered 
               according to increasing $R$ value, as indicated
               in Table \ref{ngcztab}. Except for the uppermost 
               profile ($R=110$ pix), a vertical shift was 
               applied to avoid cluttering. The shift size was 
               simply $\Delta \mu_I = n$ mag arcsec$^{-2}$, where $n$ 
               is the integer listed in the 5$^{th}$ column of the 
               mentioned table. The dotted (dashed) line is the fitted 
               exponential thin (thick) disc. The total thin+thick disc 
               contribution is given by the solid lines.}
              \label{ngc705z}%
    \end{figure}

   NGC 705 is an early type galaxy in Abell 0262, situated just 1.1
   arcmin (projected distance: $\simeq 0.03$ Mpc) away from the 
   CD galaxy NGC 708. Its redshift is the same as was assigned to  
   the cluster, $z=0.016$. Despite
   being an early-type galaxy, it has a noticeable absorption lane
   on the major axis. Its disc also displays an S-shaped twist, which
   is visible in Figure \ref{soiimgs}. NGC 705 is close enough to 
   NGC 708 that its surface photometry is contaminated by the cD's light on the
   eastern side. The contaminated region corresponds to panel {\it c}
   and, to a lesser degree, panel {\it d} in 
   Figures \ref{ngc705r} and \ref{ngc705z},
   where the horizontal and vertical profiles are shown. 
   Notice that both the $R$ and $z$ profiles in panel {\it c}
   level off at intensity levels $\simeq 0.5$
   brighter than in the other panels.
   Not surprinsingly, the regions more affected by NGC 708 are those 
   where the profile fits
   yield the flatter light distributions, specially for the thick disc
   (see Tables \ref{ngcrtab} and \ref{ngcztab}).

   Another limitation to the current analysis was that several of 
   the profiles in panel {\it b} had 
   to be shortenned due to
   contamination by the smaller early type galaxy RSCG 15. As the 
   available range for fitting the thick disc was too narrow, no fit was
   carried out in these cases. They are indicated as a ``not fit'' in
   column 3 of Table \ref{ngcztab}.

   The remaining and uncontaminated $R$ profiles are very well fit to
   a single projected exponential, with no clear truncation or
   downturn. The inner radii used are again large in order to avoid the
   bright central bulge. The average scalelength for NGC 705 is 
   $R_0 = 4.5 \pm 1.3$
   kpc when all 10 available profiles are used. However, two
   of the profiles lie beyond $2\sigma$ from this mean value. They
   are located on the eastern side of NGC 705's disc, which
   as just mentioned, is only contaminated by the outer 
   regions of the cD galaxy NGC 708. Applying a $2\sigma$ clipping here
   again substantially reduces the associated uncertainty, with
   only a mild change in the parameter itself: $R_0 = 3.9 \pm 0.5$.
   
   As in the case of ESO243G49, the fitted $R_0$ is larger away from the 
   disc plane than close to it. If we again interpret this trend as the effect of 
   a longer thick disc, we obtain $R_{02} / R_{01} \simeq 1.4$, but with
   a large scatter.

   \begin{table}
      \caption[]{NGC705 $R$ profile regions.}
         \label{ngcrtab}
     $$
         \begin{array}{p{0.30\linewidth}p{0.13\linewidth}p{0.22\linewidth}p{0.07\linewidth}p{0.08\linewidth}p{0.08\linewidth}p{0.08\linewidth}}
            \hline
            \noalign{\smallskip}
         Extract. Range & Central pixel & Fit Range & panel & n & $R_0$ & $R_0$ \\
         ($z_{min}$:$z_{max}$,$R_{min}$:$R_{max}$) & & (min:max) & & & pix & $h_{75}^{-1}$ kpc \\
            \noalign{\smallskip}
            \hline
            \noalign{\smallskip}
(-4:+4,100:230) & z=0 & (100:230) & a,b & 0 & 67.1 & 3.4 \\
(-19:-5,100:230) & z=-12.0 & (100:230) & a & 1  & 67.1 & 3.4 \\
(-44:-20,100:230) & z=-32.0 & (100:230) & a & 2 & 71.1 & 3.6 \\
(+5:+19,100:230) & z=+12.0 & (100:230) & b & 1 & 79.0 & 4.0 \\
(20:+44,100:230) & z=+32.0 & (100:230) & b & 2 & 88.9 & 4.5 \\
(-4:+4,-230:-100) & z=0 & (-230:-100) & c,d & 0 & 75.0 & 3.8 \\
(-19:-5,-230:-100) & z=-12.0 & (-230:-100) & c & 1 & 94.8 & 4.8 \\
(-44:-20,-230:-100) & z=-32.0 & (-230:-100) & c & 2 & 150.1 & 7.6 \\
(+5:+19,-230:-100) & z=+12.0 & (-230:-100) & d & 1 & 79.0 & 4.0 \\
(20:+44,-230:-100) & z=+32.0 & (-230:-100) & d & 2 & 110.6 & 5.6 \\
            \noalign{\smallskip}
            \hline
         \end{array}
     $$
\begin{list}{}{}
\item [] Notes: Column 1: Range perpendicular (zmin:zmax) and parallel 
(Rmin:Rmax) to major axis where the profile was extracted, in pixels. Pixel
scale is 0.10arcsec/pixel; column 2: central pixel position, in pixels; 
column 3: fit range in pixels used; column 4: panel in 
Figure \ref{ngc705r} where the 
profile is shown; column 5: artificial shift in surface brightness 
applied to profile in its panel, $\Delta \mu_r = n$ mag arcsec$^2$.
The larger $n$, the lower the profile position in its panel; column 6:
fitted scalength, $R_0$, in pixels; column 7: scalelength in
units of $h_{75}^{-1}~kpc$, where $h_{75} = H_0 / 75$ and 
$H_0$ is the Hubble constant in units of km s$^{-1}$ Mpc$^{-1}$.
\end{list}
   \end{table}

   As for the $z$ profiles, the need for a double
   exponential is again visible, as most of the profiles are not straight
   lines in Figure \ref{ngc705z}. However, a few profiles
   extracted at higher $R$ are consistent with a single
   exponential. In fact, attempts to fit a double exponential failed for
   them. Their single scaleheights are listed within parenthesis in Table 
   \ref{ngcztab}. They are usually intermediate between the thin and thick
   disc scaleheights obtained from the other profiles, attesting to the
   increasing difficulty of differentiating the two components at faint
   $\mu_I$ levels.

   The average thin disc scaleheight
   is $z_{01} = 0.5 \pm 0.15$ kpc, where the uncertainty comfortably
   accomodates all individual profile fits. Notice that this is a
   relatively nearby galaxy (as compared to ESO187G19 and LCSBS0946P),
   which is imaged with HST; the effects of seeing are therefore
   minimized. Thus, its smaller thin disc scaleheight may indicate
   that the $z_{01}$ values from the SOAR/SOI images are only 
   upper limits. The value of $z_{01}$ for NGC 705 is 
   comparable to that of the Galaxy or other nearby spirals.

   The thick disc of NGC 705 also has a lower derived scaleheight 
   than those from the SOAR/SOI galaxies. Eliminating the
   two clear outliers in Table \ref{ngcztab} leads to $z_{02} = 2.3 \pm 
   1.2$ kpc. As it is impossible that seeing would have a strong effect
   on an extended thick disc structure, we conclude  
   that NGC 705 in fact has a generally smaller disc structure when
   compared to the other galaxies.

   \begin{table}
      \caption[]{NGC705 $z$ profile regions.}
         \label{ngcztab}
     $$
         \begin{array}{p{0.28\linewidth}p{0.13\linewidth}p{0.26\linewidth}p{0.05\linewidth}p{0.04\linewidth}p{0.08\linewidth}p{0.08\linewidth}}
            \hline
            \noalign{\smallskip}
         Extract. Range & Central pixel & Fit Range & pnl & n & $z_{01}$, $z_{02}$ & $z_{01}$, $z_{02}$ \\
         ($z_{min}$:$z_{max}$,$R_{min}$:$R_{max}$) & & (min:max) & & & pix & $h_{75}^{-1}$ kpc \\
            \noalign{\smallskip}
            \hline
            \noalign{\smallskip}
(-105:-1,100:120) & R=110 & Thin:(-70:-20), Thick:(-100:-75) & a & 0 & 9.7, 35.5 & 0.3, 1.1 \\
(1:66,100:120) & R=110 & No fit & b & 0 & NA, NA & NA, NA \\ 
(-105:-1,120:144) & R=132 & Thin:(-70:-20), Thick:(-100:-75) & a & 1 & 11.9, 38.4 & 0.4, 1.2 \\
(1:66,120:144) & R=132 & No fit & b & 1 & NA, NA & NA, NA \\
(-105:-1,145:180) & R=162.5 & Thin:(-70:-20), Thick:(-100:-75) & a & 2 & (28.3) & (0.9) \\
(1:66,145:180) & R=162.5 & No fit & b & 2 & NA, NA & NA, NA \\
(-105:-1,180:225) & R=202.5 & Thin:(-70:-20), Thick:(-100:-75) & a & 3 & (25.4) & (0.8) \\
(1:87,180:225) & R=202.5 & Thin:(20:70), Thick:(75:87) & b & 3 & 12.4, 65.7 & 0.4, 2.1 \\
(-105:-1,225:270) & R=247.5 & Thin:(-70:-20), Thick:(-100:-75) & a & 4 & (25.0) & (0.8) \\
(1:105,225:270) & R=247.5 & Thin:(20:70), Thick:(75:100) & b & 4 & (34.5) & (1.1) \\
(-96:-1,-120:-100) & R=-110 & Thin:(-70:-20), Thick:(-96:-75) & c & 0 & 18.1, 101.6 & 0.6, 3.2 \\
(1:105,-120:-100) & R=-110 & Thin:(20:70), Thick:(75:100) & d & 0 & 14.2, 51.0 & 0.4, 1.6 \\ 
(-96:-1,-144:-120) & R=-132 & Thin:(-70:-20), Thick:(-96:-75) & c & 1 & NA, NA & NA, NA \\
(1:105,-144:-120) & R=-132 & Thin:(20:70), Thick:(75:100) & d & 1 & 15.3, 59.9 & 0.5, 1.9 \\
(-96:-1,-180:-145) & R=-162.5 & Thin:(-70:-20), Thick:(-96:-75) & c & 2 & NA, NA & NA, NA \\
(1:105,-180:-145) & R=-162.5 & Thin:(20:70), Thick:(75:100) & d & 2 & 21.6, 155.0 & 0.7, 4.9 \\
(-96:-1,-225:-180) & R=-202.5 & Thin:(-70:-20), Thick:(-96:-75) & c & 3 & 23.2, 804.0 & 0.7, 25.4 \\
(1:105,-225:-180) & R=-202.5 & Thin:(20:70), Thick:(75:100) & d & 3 & 12.8, 74.9 & 0.4, 2.4 \\
(-96:-1,-270:-225) & R=-247.5 & Thin:(-70:-20), Thick:(-96:-75) & c & 4 & 17.9, 427.6 & 0.6, 13.5 \\
(1:105,-270:-225) & R=-247.5 & Thin:(20:70), Thick:(75:100) & d & 4 & 9.5, 74.1 & 0.3, 2.3 \\
            \noalign{\smallskip}
            \hline
         \end{array}
     $$
\begin{list}{}{}
\item [] Notes: Column 1: Range perpendicular (zmin:zmax) and parallel 
(Rmin:Rmax) to major axis where the profile was extracted, in pixels. Pixel
scale is 0.10arcsec/pixel; column 2: central pixel position, in pixels; 
column 3: fit range, in pixels, used for the thin and thick components; 
column 4: panel in Figure \ref{ngc705z} where the 
profile is shown; column 5: artificial shift in surface brightness 
applied to profile in its panel, $\Delta \mu_r = n$ mag arcsec$^2$.
The larger $n$, the lower the profile position in its panel; column 6:
fitted scaleheights, $z_{01}$ and $z_{02}$, in pixels; column 7: 
scaleheights in units of $h_{75}^{-1}~kpc$, where $h_{75} = H_0 / 75$ and 
$H_0$ is the Hubble constant in units of km s$^{-1}$ Mpc$^{-1}$.
\end{list}
   \end{table}

\subsection{Surface brightnesses}

So far our analysis has been limited to measuring the horizontal and vertical
scales of both thin and thick disc components in the 4 galaxies. In this 
section we turn our attention to the expected surface brightness of the
two components along the major axis and at their centre. We attempt to 
reconstruct the major axis profiles by extrapolating the fitted $z$ profiles
all the way to the disc plane. In doing so, as long as the extrapolations 
are valid and the fits avoid the areas under dust extinction, we should
be able to recover the mid-plane surface brightness of both thin and thick 
discs at the different $R$ values where the $z$ profiles were extracted.

   \begin{figure}
   \centering
   \includegraphics[width=0.5\textwidth]{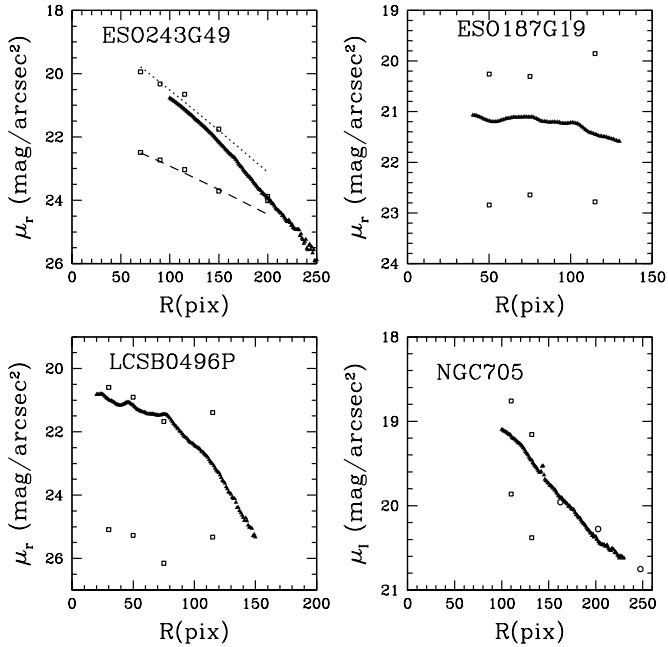}
   \caption{Squares: Surface brightness values of thin and thick discs derived 
               by extrapolating the fitted $z$ profiles towards the disc 
               mid-plane. Triangles: observed major axis profiles. 
               In the upper left panel (ESO243G49) the dotted and dashed lines
               are fits to 5 points in the extrapolated mid-plane profile.}
              \label{majoraxis}%
    \end{figure}

In Figure \ref{majoraxis} we show the results of applying this exercise 
to the 4 galaxies. The squares in the panels represent the thin and thick
disc $\mu (R,z=0)$ values as a function of $R$. They were obtained
by fitting a double exponential to the mean profile among the different 
quadrants at a fixed $R$ and extrapolating the fit result to $z=0$.
The triangles show the average of the 2 observed profiles along the major 
axis, one on each
side of the centre of the galaxy. For LCSBS0496P and NGC705, only one major
axis profile was used, since the other one was contaminated by scattered light
and by the cluster cD galaxy, respectively. 

For ESO243G49, we obtained $\mu_r (R,0)$ for all 5 $R$ positions studied. 
We then fitted Eq. (1) to these values (shown as dotted and dashed lines,
respectively, for the thin and thick components). Extrapolating all the way
towards the centre, we infer $\mu_{r,1} (0,0) \simeq 18.5$ mag arcsec$^{-2}$
for the thin disc and $\mu_{r,2} (0,0) \simeq 21.5$ mag arcsec$^{-2}$ 
for the thick disc. 
Thus, at the centre, the thick disc roughly corresponds to $6\%$ of the
light contribution from the thin disc. Due to the difference in
scalelengths, at the down-bending radius ($R \simeq 160$ pix), 
this fraction increases to about $25\%$. We also estimate
the dust extinction coefficient at the disc plane of ESO243G49 as 
$A_r \simeq 0.3-0.5$ mag,
based on the comparison between the predicted and observed major axis profiles.

For the other galaxies, substantially less information is available for 
different reasons. We only have 3 $R$ points along the ESO187G19 major
axis. For all of them we managed to recover a $\mu_r (R,0)$ value 
from the 
average $z$ profiles. The amount of extinction on the disc plane is
much greater, as expected, than in ESO243G49: $A_r \simeq 1.0$ mag. Both
thin and thick discs follow the observed flat projected profile along the 
major axis. The thick/thin disc brightness ratio at $z=0$ 
is $\simeq 0.1 = 10\%$ in the 3 positions along the major axis.

LCSBS0496P has a very faint thick disc, some 4 magnitudes fainter than
the thin disc. This corresponds to a relative normalization along
the major axis of $2\%$, which is comparable to or smaller than the
Galactic thick disc normalization in the solar neighbourhood.
A visual extrapolation towards $R=0$ leads to a central surface brightnesses
of $\mu_{r,1} \simeq 20$ mag arcsec$^{-2}$ and $\mu_{r,2} \simeq 24.5$ 
mag arcsec$^{-2}$.
The inferred amount of extinction is $A_r \simeq 0.3$ mag, but with large
scatter.

Finally, for NGC 705, we used the profiles in panel {\it a} of  
Figure \ref{ngc705z} to find $\mu (R,0)$. The other ones were
contaminated by light from neighbouring galaxies. Only two
successful fits to both the thin and thick discs are available for the 
quadrant we used. They show a thick disc about $1.2$ mag fainter than
the thin disc, contributing therefore with about $30\%$ of the z=0
light at an intermediate range along the major axis. We also plot the 
results of extrapolating the profile fits using a single disc component, 
whose results are shown in parenthesis in Table \ref{ngcztab}. They are 
shown as open circles and follow the observed major axis 
profile more closely. We thus conclude that dust is very concentrated 
towards the centre of NGC 705, yielding an absorption of $A_I \simeq 
0.4$ mag within $R \simeq 140$ pix $=4.4$ kpc.

\section{Discussion and conclusions}

   We carried out deep surface photometry of 4 edge-on spiral
   galaxies located in Abell clusters. Three of them were imaged with 
   the SOAR/SOI detector and one was taken from the HST/WFPC2 archive.
   The surface brightness levels reached are $\mu_{lim} \simeq 26-26.5$
   mag arcsec$^{-2}$.

   The profiles parallel to the major axis of two galaxies,
   ESO243G49 and LCSBS0496P, do not follow a single projected exponential
   function. They display a downturn at outer radii. This has been
   previously described by several authors; see, e.g., the review by 
   Pohlen et al. (2004b). According to their $R$ profile shapes, 
   these two galaxies fit into the Type II profile class in the
   scheme originally proposed by Freeman (1970) and later extended 
   by Erwin, Beckman \& Pohlen (2005) and Pohlen \& Trujillo (2006).
   The 2 other galaxies, ESO187G19 and NGC 705 exhibit horizontal
   profiles that are well fit by a single exponential over the
   entire range studied (Type I profiles according to the same scheme).

   Profile fits to single exponential models in directions 
   parallel to the major axis have resulted in scalelengths ($R_0$) for
   each galaxy. In the case of the Type II profiles mentioned
   earlier, the fits were restricted to the region inside the truncation
   radius. The two early-type galaxies, ESO243G49 and NGC705, 
   have lower $R_0$ values
   close to the disc plane than at larger distances from it. If we 
   interpret this trend as a difference in the horizontal structure of 
   thin and thick
   discs, we derive a ratio $R_{02} / R_{01} \simeq 1.4$, although with a
   significant scatter from one fit to another. This is smaller than
   but still consistent with the mean thick/thin disc scalelength 
   ratio of 1.9 quoted by Pohlen et al. (2004) for their sample of 
   S0 galaxies. We find no significant trend in $R_0$ for our two 
   late-types, ESO187G19 and LCSBS0496P. Results from
   the literature also suggest a similar dependence of $R_{02} / R_{01}$ 
   on morphological type. For the Milky-Way and NGC 4565, this ratio 
   is around
   1.1-1.4 (Shaw \& Gilmore 1989; Ojha 2001; Wu et al. 2002; 
   Larsen \& Humphreys 2003), whereas for the late
   type ESO 342G17, Neeser et al. (2002) have found a ratio close 
   to unity.

   As for the physical values of $R_0$ we obtained, they are of course 
   dependent on the adopted distance scale. We here express them in
   units of $h_{75}$, which is the Hubble constant in units of
   $75$ km s$^{-1}$ Mpc$^{-1}$. The values vary considerably across 
   the sample ($3.5 < R_0 < 12.9$ kpc, where $R_0$ is the mean over all
   profiles). These values of $R_0$ are typical. Pohlen et al. (2004) list 
   $2.5 < R_{02} < 16.7$ kpc and $1.3 < R_{01} < 9.6$ kpc for their sample 
   of 6 lenticular galaxies . For the Galaxy, these values are 
   $R_{01} \simeq 3.0$ kpc and $R_{02} \simeq 4.2$ kpc as quoted by 
   Ojha (2001) and Larsen \& Humphreys (2003).

   We also analysed the light distribution perpendicular to the disc planes.
   As a rule, in all galaxies the extracted vertical profiles are inconsistent
   with a single exponential, requiring a two-component fit. Our derived
   thick-disc / thin-disc scaleheight ratios are $z_{02} / z_{01} \simeq 4$
   for 3 of the galaxies and twice this value for LCSBS0496P. This is quite
   similar to values quoted in the literature (see compilation in Pohlen et 
   al 2007). In terms of physical units, our results also fit well into
   the range quoted by previous studies: $2.3 < z_{02} < 6.2$ kpc for
   the thick disc, and $0.5 < z_{02} < 1.2$ kpc for the thin disc. 
   Given the large distance of 
   ESO187G19 and LCSBS0496P ($z \simeq 0.04-0.05$), however, the 
   scaleheights
   we obtain for the thin disc must be considered with caution, since they
   are comparable to the photometric seeing. To better analyse 
   these and other galaxies in high-density regions, we are currently 
   developing a modelling algorithm that will deconvolve the observed
   images with the point spread function and recover the 3D figure of 
   edge-on spiral galaxies.
   
   By extrapolating the fitted vertical profiles, we derived estimates
   of the surface brighness values for both disc
   components at different positions along the major axis of each galaxy.
   From the rough $\mu (R,0)$ profiles, we could estimate the thick/thin
   disc normalization at different $R$ values. The fractional contribution
   of the thick component relative to the thin disc varies from $\simeq 2\%$ 
   at the mid-plane of LCSBS0496B to $\simeq 30\%$ at intermediate major
   axis positions in NGC 705. For ESO243G49 and ESO187G19, the thick disc
   typically contributes with $\simeq 10\%$ of the light. Estimates
   of the amount of dust extinction along the major axis were also obtained 
   by comparing the
   expected surface brightness at the mid-plane to the observed ones.

   In brief, we conclude that the structural parameters for both planar 
   components in our
   cluster galaxy sample are similar to those typically found in the
   general field. We stress that 3 of our sample galaxies are very
   close to their host cluster centre, and should therefore have suffered
   dynamical effects, such as interactions, accretions, or harassment. 
   Taken at face value, although with a small sample, our results imply
   that these dynamical effects typical of high density regions may
   have a limited influence on the sizes of the thin and thick discs.

\begin{acknowledgements}
      This work was supported by Conselho Nacional de Desenvolvimento
      Cient\'\i fico e Tecnol\'ogico (CNPq) in Brazil through a research
      grant to BXS and a doctoral fellowship to TBV. The authors thank
      the staff at SOAR for collecting the data.
\end{acknowledgements}

\end{document}